\documentclass[twocolumn,english,pra, aps, twocolumn, superscriptaddress]{revtex4-1}
\usepackage[T1]{fontenc}
\usepackage[latin9]{inputenc}
\usepackage{color}
\usepackage{mathrsfs}
\usepackage{amsmath}
\usepackage{amssymb}
\usepackage{graphicx}

\makeatletter
\usepackage{xcolor}
\usepackage[colorlinks,citecolor=blue,linkcolor=red]{hyperref}

\makeatother

\usepackage{babel}
\begin{document}
\title{Works with quantum resource of coherence}
\author{Yu-Han Ma}
\address{Graduate School of China Academy of Engineering Physics, No. 10 Xibeiwang
East Road, Haidian District, Beijing, 100193, China}
\author{C. L. Liu}
\address{Graduate School of China Academy of Engineering Physics, No. 10 Xibeiwang
East Road, Haidian District, Beijing, 100193, China}
\author{C. P. Sun}
\email{suncp@gscaep.ac.cn}

\address{Graduate School of China Academy of Engineering Physics, No. 10 Xibeiwang
East Road, Haidian District, Beijing, 100193, China}
\address{Beijing Computational Science Research Center, Beijing 100193, China}
\begin{abstract}
We study the modification of the second law of thermodynamics for
a quantum system interacting with a reservoir regarding quantum coherence.
The whole system is isolated so that neither energy nor information
is lost. It is discovered that the coherence of the reservoir can
serves as a useful resource allowing the system extract more energy
from the reservoir; \textcolor{black}{among the coherence measures,
only is the relative entropy of coherence feasible to }quantitatively\textcolor{black}{{}
characterize energy exchange.} We demonstrate that a thermodynamic
cycle between two coherent reservoirs can output more work than its
classical counterpart. The efficiency of such cycle surpasses the
Carnot efficiency, which is the upper bound of heat engine efficiency
in classical regime. 
\end{abstract}
\maketitle
\textit{Introduction.-}Conventional thermodynamics deals with macro-matters
in equilibrium states. The thermodynamic limit, introduced as a mathematical
approach, restricts this theoretical framework to macroscopic situation
\citep{Carnoteff}. When dealing with the thermodynamic properties
and energy conversion for microscopic systems off thermodynamic limit,
it is necessary to take into account quantum coherence effects \citep{Esposito_2009,Campisi_2011,goold2016role,streltsov2017colloquium}.
This goes beyond the valid regime of the conventional thermodynamics.
Over the past few decades, quantum thermodynamics was developed \citep{Vinjanampathy_2016,Binder2018,kosloff2019quantum,su2021thermodynamic}
to build a bridge between the micro-world dominated by quantum mechanics
and the macro-world by classical thermodynamics. 

Recently, quantum resource theories were proposed to quantitatively
describe the usefulness of the quantum coherence effects \citep{brandao2013resource,gour2015resource,goold2016role,sparaciari2017resource,streltsov2017colloquium,lostaglio2019introductory}.
For thermodynamics, an immediate question raises: how quantum coherence
effects enhance the energy conversion for non-equilibrium processes
in quantitative level \citep{alicki2013entanglement,campisi2016power,YHMaQPTHE,bresque2021two}.
Recent studies have shown that quantum coherence resource is a promising
candidate for such enhancement role \citep{scully2003extracting,quan2006quantum,scully2011quantum,park2013heat,NanoscaleHEPRL2014,uzdin2015equivalence,dorfman2018efficiency,Camati2019,su2021coherence}.
However, how to characterize the role of quantum coherence in thermodynamics
with a general approach remains an open problem, although extensive
attempts have been made for some specific systems \citep{lostaglio2015description,kammerlander2016coherence,korzekwa2016extraction,Su2018,francica2019role,rodrigues2019thermodynamics,de2020unraveling}.
This challenge is not only essential to the foundations of quantum
thermodynamics, but also crucial for designing various energy conservation
machines operating in the micro-world with high-performance. 

In this Letter, we revisit the second law of thermodynamics for a
quantum system interacting with a reservoir. Both of the system and
reservoir are of finite size so that the quantum coherence can persist
in a duration in comparison with the relaxation time and the cycle
time. We derive the heat exchange and entropy flow inside such an
interacting system which is not thermally equilibrium. The obtained
result shows that the second law of thermodynamics is quantitatively
modified with the quantum coherence resource, which is characterized
by the relative entropy of coherence $\mathscr{C}$ \citep{Quantifying2014}.
The modified second law of thermodynamics in quantum regime is further
applied to a non-equilibrium thermodynamic process with a finite-sized
reservoir. It is discovered that the quantum coherence of the reservoir
is converted into heat, which can be extracted as exceeding work.
When a quantum Carnot heat engine operates between two coherent reservoirs,
the corresponding efficiency will surpass its classical counterpart,
namely, the Carnot efficiency.

\textit{Thermodynamics in an isolated binary system.-}The isolated
system $\mathrm{SE}$ of interest is composed of the system $\mathrm{S}$
and the finite-sized reservoir $\mathrm{E}$. The conventional thermodynamics
regards $\mathrm{E}$ as an infinite system. The total Hamiltonian
of such coupling system is driven by a time-dependent parameter $\lambda\left(t\right)$
as

\begin{equation}
H\left(t\right)=H_{\mathrm{S}}\left[\lambda\left(t\right)\right]+H_{\mathrm{E}}+V.
\end{equation}
Here $H_{\mathrm{S}}\left[\lambda\left(t\right)\right]$ and $H_{\mathrm{E}}$
are the Hamiltonian of $\mathrm{S}$ and $\mathrm{E}$, respectively;
$V$ describes the interaction between $\mathrm{S}$ and $\mathrm{E}$,
which will result \textcolor{black}{in entropy exchange and heat exchange
between }$\mathrm{S}$ and $\mathrm{E}$\textcolor{black}{{} during
the evolution of $\mathrm{SE}$. Work can be applied to the system
$\mathrm{S}$ by tuning $\lambda\left(t\right)$. Hereafter, for simplicity,
we consider the }reservoir\textcolor{black}{{} is non-degenerate in
energy, and the \textquotedblleft coherence\textquotedblright{} of
the }reservoir\textcolor{black}{{} is defined with respect to energy
eigen-states of $\mathrm{E}$. Assuming the whole system $\mathrm{SE}$
is initially prepared in the product state $\rho_{\mathrm{SE}}(0)=\rho_{\mathrm{S}}(0)\otimes\rho_{\mathrm{E}}(0)$}
with $\rho_{\alpha}$ ($\alpha=\mathrm{SE,S,E}$) the density matrix
of $\alpha$. Such a product state $\rho_{\mathrm{SE}}(0)$ implies
there is no initial correlation between $\mathrm{S}$ and $\mathrm{E}$.
The state of $\mathrm{SE}$ subjects to unitary evolution at time
$t$ is $\rho_{\mathrm{SE}}(t)=U(t)\rho_{\mathrm{SE}}(0)U^{\dagger}(t)$,
where $U(t)=\mathscr{\mathcal{T}}\exp\left[-i/\hbar\int_{0}^{t}H(t)dt'\right]$
is the evolution operator and $\mathscr{\mathcal{T}}$ is the time-order
operation. The time evolution of the sub-system $\mathrm{S}$ ($\mathrm{E}$)
is described by the reduced density matrix $\rho_{\mathrm{S(E)}}\left(t\right)=\mathrm{Tr}_{\mathrm{E(S)}}\left[\rho_{\mathrm{SE}}\left(t\right)\right]$.
It is stressed here that in most studies, the reservoir is considered
as a macro-system with infinite size, so that its thermal equilibrium
state will not change over certain time, namely, $\rho_{\mathrm{E}}(0)\rightarrow\rho_{\mathrm{E}}(t)=\rho_{\mathrm{E}}(0)=\exp(-\beta_{\mathrm{E}}H_{\mathrm{E}})/\mathrm{Tr}_{\mathrm{E}}\left[\exp(-\beta_{\mathrm{E}}H_{\mathrm{E}})\right]$.
Here $\beta_{\mathrm{E}}=1/(k_{\mathrm{B}}T_{\mathrm{E}})$ is the
inverse temperature of $\mathrm{E}$, $T_{\mathrm{E}}$ the temperature
of $\mathrm{E}$, and $k_{\mathrm{B}}$ the Boltzmann constant ($k_{\mathrm{B}}=1$
is used in the following discussion). 

To modify the second law of thermodynamics in quantum regime, we first
consider the first law of thermodynamics for $\mathrm{S}$. Since
the whole system $\mathrm{SE}$ is isolated, the change in internal
energy of $\mathrm{SE}$, namely, $E_{\mathrm{SE}}(t)\equiv\left\langle H(t)\right\rangle =\mathrm{Tr}\left[\rho_{\mathrm{SE}}(t)H(t)\right]$,
is achieved through the outputting work or applied work, that is

\begin{equation}
\dot{E}_{\mathrm{SE}}(t)=\dot{\lambda}\mathrm{Tr}\left[\rho_{\mathrm{S}}(t)\frac{\partial H_{\mathrm{S}}}{\partial\lambda}\right]\equiv\dot{W}_{\mathrm{S}}(t),\label{eq:Ws}
\end{equation}
where, for any thermodynamic variable $\Lambda$, $\dot{\Lambda}=d\Lambda/dt$.
With the weak coupling approximation \citep{rivas2020strong}, namely,
the interaction $V$ is small enough in comparison with $H_{\mathrm{S}}$
and $H_{\mathrm{E}}$, one has

\begin{equation}
\dot{E}_{\mathrm{SE}}(t)\mathcal{\approx}\frac{d}{dt}\left\langle H_{\mathrm{S}}\left[\lambda(t)\right]\right\rangle +\frac{d}{dt}\left\langle H_{\mathrm{E}}\right\rangle \label{eq:EUdot}
\end{equation}
Here, the change in the internal energy of $\mathrm{E}$, i.e., $d\left\langle H_{\mathrm{E}}\right\rangle /dt$,
is defined as the heat exchange between $\mathrm{S}$ and $\mathrm{E}$,
i.e., $\mathrm{Tr}\left[\dot{\rho}_{\mathrm{E}}(t)H_{\mathrm{E}}\right]\equiv\dot{Q}_{\mathrm{E}}(t)=-\dot{Q}_{\mathrm{S}}(t)$.
It follows from Eqs. (\ref{eq:Ws}) and (\ref{eq:EUdot}) that 

\begin{equation}
\dot{E}_{\mathrm{S}}(t)=\frac{d}{dt}\left\langle H_{\mathrm{S}}\left[\lambda(t)\right]\right\rangle =\dot{W}_{\mathrm{S}}(t)+\dot{Q}_{\mathrm{S}}(t).\label{eq:1st-law}
\end{equation}
With the obtained first law of thermodynamics in Eq. (\ref{eq:1st-law}),
we go further to study the second law of thermodynamics in this scenario.
Generally, as the entire system $\mathrm{SE}$ evolves, $\mathrm{S}$
and $\mathrm{E}$ will become entangled, resulting in correlation
between $\mathrm{S}$ and $\mathrm{E}$. Such correlation can be measured
by the mutual information $I=I(\mathrm{S}:\mathrm{E})$ as \citep{nielsen2002quantum,li2017production}

\begin{equation}
I(t)\equiv S_{\mathrm{S}}(t)+S_{\mathrm{E}}(t)-S_{\mathrm{SE}}(t),
\end{equation}
with $S_{\mathrm{\alpha}}(t)\equiv\mathrm{-Tr}\left[\rho_{\alpha}(t)\mathrm{ln}\rho_{\alpha}(t)\right]$
($\alpha=\mathrm{SE,S,E}$) being the von Neumann entropy of $\alpha$
at time $t$. Taking the time derivative of the above equation, we
can express entropy flow of the system as

\begin{equation}
\dot{S}_{\mathrm{S}}(t)=\dot{I}(t)-\dot{S}_{\mathrm{E}}(t),\label{eq:Ssdot}
\end{equation}
where the relation $\dot{S}_{\mathrm{SE}}\left(t\right)=0$ is used
since the evolution of $\mathrm{SE}$ is unitary. The entropy flow
of $\mathrm{E}$, i.e., $\dot{S}_{\mathrm{E}}(t)$, can be re-written
with the heat flow $\dot{Q}_{\mathrm{S}}$ as (See Supplementary Materials
(SM) for detailed derivation \citep{SM1})

\begin{equation}
\dot{S}_{\mathrm{E}}=-\dot{S}\left[\rho_{\mathrm{E}}(t)||\rho_{\mathrm{E}}^{\mathrm{eq}}(0)\right]-\beta_{\mathrm{E}}\dot{Q}_{\mathrm{S}}.\label{eq:SEdot}
\end{equation}
Here, $S\left[\rho_{\mathrm{E}}(t)||\rho_{\mathrm{E}}^{\mathrm{eq}}(0)\right]\equiv\mathrm{tr}\left[\rho_{\mathrm{E}}(t)\mathrm{ln}\rho_{\mathrm{E}}(t)\right]-\mathrm{tr}\left[\rho_{\mathrm{E}}(t)\mathrm{ln}\rho_{\mathrm{E}}^{\mathrm{eq}}(0)\right]$
is the quantum relative entropy between the state $\rho_{\mathrm{E}}(t)$
and $\rho_{\mathrm{E}}^{\mathrm{eq}}(0)$; $\rho_{\mathrm{E}}^{\mathrm{eq}}(0)\equiv e^{-\beta_{\mathrm{E}}H_{\mathrm{E}}}/\mathrm{tr}\left(e^{-\beta_{\mathrm{E}}H_{\mathrm{E}}}\right)$
is the initial effective equilibrium density matrix of the reservoir
with $\beta_{\mathrm{E}}=\beta_{\mathrm{E}}(0)$ the corresponding
effective initial inverse temperature of $\mathrm{E}$. It follows
from Eqs. (\ref{eq:Ssdot}) and (\ref{eq:SEdot}) that 

\begin{equation}
\Delta S_{\mathrm{S}}=\beta_{\mathrm{E}}\Delta Q_{\mathrm{S}}+\Delta I+\Delta S\left[\rho_{\mathrm{E}}(t)||\rho_{\mathrm{E}}^{\mathrm{eq}}(0)\right],\label{eq:DS-total}
\end{equation}
where $\Delta(\bullet)\equiv\int_{0}^{t}\dot{\text{(\ensuremath{\bullet})}}dt=\bullet(t)-\bullet(0).$
The second and third terms in the right hand of the above equation
relate to the irreversible entropy generation $\Delta S_{\mathrm{S}}^{(\mathrm{ir})}=\Delta S_{\mathrm{S}}-\beta_{\mathrm{E}}\Delta Q_{\mathrm{S}}$
\citep{spohn1978entropy,li2017production,Constraintrelationyhma,2020IEGyhma}.
Notice that the mutual information is non-negative, namely, $I(t)\geq0$
\citep{nielsen2002quantum}, and there is no initial correlation between
the system and reservoir, i.e., $I(0)=0$, then $\Delta I\geq0$.
Therefore, the second law of thermodynamics is obtained as

\begin{equation}
\Delta S_{\mathrm{S}}\geq\frac{\Delta Q_{\mathrm{S}}}{T_{\mathrm{E}}}+\Delta S\left[\rho_{\mathrm{E}}(t)||\rho_{\mathrm{E}}^{\mathrm{eq}}(0)\right],\label{eq:2nd-quantum}
\end{equation}
where $T_{\mathrm{E}}=\beta_{\mathrm{E}}^{-1}$ is the initial effective
temperature of the reservoir. 

As we mentioned before, in the usual treatment, the reservoir is considered
as a classical infinite thermal reservoir, thus the state of $\mathrm{E}$
will not be affected by $\mathrm{S}$ and will remain unchanged during
the evolution of $\mathrm{SE}$, namely, $\rho_{\mathrm{E}}(t)=\rho_{\mathrm{E}}^{\mathrm{eq}}(0)$.
In this sense, $\Delta S\left[\rho_{\mathrm{E}}(t)||\rho_{\mathrm{E}}^{\mathrm{eq}}(0)\right]=0$,
and then Eq. (\ref{eq:2nd-quantum}) reduces to the familiar form
$\Delta S_{\mathrm{S}}\geq\beta_{\mathrm{E}}\Delta Q_{\mathrm{S}}$,
which is the second law of thermodynamics in classical regime. In
following discussion, we will analyze the second term of Eq. (\ref{eq:2nd-quantum}),
i.e., $\Delta S\left[\rho_{\mathrm{E}}(t)||\rho_{\mathrm{E}}^{\mathrm{eq}}(0)\right]$,
in detail. The role of the reservoir's finite size and coherence in
the modified second law of thermodynamics will be distinguished and
clearly presented.

\textit{Finite size reservoir with coherence.-}To further explore
the quantum properties of $\Delta S\left[\rho_{\mathrm{E}}(t)||\rho_{\mathrm{E}}^{\mathrm{eq}}(0)\right]$,
we first divide $\rho_{\mathrm{E}}(t)$ into the diagonal part $\rho_{\mathrm{E}}^{\mathrm{d}}(t)$
and non-diagonal part $\rho_{\mathrm{E}}^{\mathrm{nd}}(t)$, namely,
$\rho_{\mathrm{E}}(t)=\rho_{\mathrm{E}}^{\mathrm{d}}(t)+\rho_{\mathrm{E}}^{\mathrm{nd}}(t)$,
where $\rho_{\mathrm{E}}^{\mathrm{nd}}(t)$ is non-vanishing for coherent
reservoirs. Then, $S\left[\rho_{\mathrm{E}}(t)||\rho_{\mathrm{E}}^{\mathrm{eq}}(0)\right]$
can be divided into three parts as \citep{SM1}

\begin{equation}
S\left[\rho_{\mathrm{E}}(t)||\rho_{\mathrm{E}}^{\mathrm{eq}}(0)\right]=\mathscr{C}_{\mathrm{E}}(t)+S\left[\rho_{\mathrm{E}}^{\mathrm{d}}(t)||\rho_{\mathrm{E}}^{\mathrm{eq}}(0)\right]+\frac{\Delta Q_{\mathrm{S}}^{2}}{2C_{\mathrm{E}}T_{\mathrm{E}}^{2}}.\label{eq:Sroud}
\end{equation}

Here, $\mathscr{C}_{\mathrm{E}}(t)\equiv\mathrm{tr}\left[\rho_{\mathrm{E}}(t)\mathrm{ln}\rho_{\mathrm{E}}(t)-\rho_{\mathrm{E}}^{\mathrm{d}}(t)\mathrm{ln}\rho_{\mathrm{E}}^{\mathrm{d}}(t)\right]$
is the \textit{relative entropy of coherence, }which is a measure
of quantum coherence in the resource theory of coherence \citep{Quantifying2014,streltsov2017colloquium,horodecki2013quantumness,brandao2015reversible,chitambar2019quantum}.
$\rho_{\mathrm{E}}^{\mathrm{eq}}(t)\equiv e^{-\beta_{\mathrm{E}}(t)H_{\mathrm{E}}}/\mathrm{tr}\left[e^{-\beta_{\mathrm{E}}(t)H_{\mathrm{E}}}\right]$
is the effective equilibrium density matrix of $\mathrm{E}$ at time
$t$, the effective inverse temperature $\beta_{\mathrm{E}}(t)$ is
determined by the internal energy $U_{\mathrm{E}}(t)$, $\Delta Q_{\mathrm{S}}=\int_{0}^{t}\dot{Q}_{\mathrm{S}}dt$
is the heat exchange between $\mathrm{S}$ and $\mathrm{E}$, and
$C_{\mathrm{E}}$ is the heat capacity of $\mathrm{E}$. Physically,
$\mathscr{C}_{\mathrm{E}}(t)$ represents the contribution of coherence,
and $S\left[\rho_{\mathrm{E}}^{\mathrm{d}}(t)||\rho_{\mathrm{E}}^{\mathrm{eq}}(t)\right]$
describes the degree to which the reservoir deviates from equilibrium.
The term $\Delta Q_{\mathrm{S}}^{2}/\left(2C_{\mathrm{E}}T_{\mathrm{E}}^{2}\right)$
is caused by the back-action of $\mathrm{S}$ to $\mathrm{E}$ under
the constraint of energy conservation \citep{reeb2015tight,richens2018finite,timpanaro2020landauer}.
Specifically, with the heat exchange between $\mathrm{S}$ and $\mathrm{E}$,
the internal energy of $\mathrm{E}$ changes, and the effective temperature
of $\mathrm{E}$ thus changes accordingly. This effect will reduce
the efficiency at maximum work of a heat engine operating between
finite-sized heat reservoirs \citep{izumida2014work,2020-finite-size}.
From Eq. (\ref{eq:DS-total}) and Eq. (\ref{eq:Sroud}), we obtain 

\begin{equation}
\Delta S_{\mathrm{S}}=\frac{\Delta Q_{\mathrm{S}}}{T_{\mathrm{E}}}+\frac{\Delta Q_{\mathrm{S}}^{2}}{2C_{\mathrm{E}}T_{\mathrm{E}}^{2}}+\Delta S\left[\rho_{\mathrm{E}}^{\mathrm{d}}(t)||\rho_{\mathrm{E}}^{\mathrm{eq}}(t)\right]+\Delta\mathscr{C}_{\mathrm{E}}+\Delta I,\label{eq:Mainresult}
\end{equation}
where $\Delta\mathscr{C}_{\mathrm{E}}=\mathscr{C}_{\mathrm{E}}(t)-\mathscr{C}_{\mathrm{E}}(0)$
is the change in coherence of the reservoir during the whole process.
In order to compare the current result with that of the conventional
thermodynamics, we further assume that the diagonal elements the reservoir's
density matrix follows the time-dependent Boltzmann distribution,
namely, $\rho_{\mathrm{E}}^{\mathrm{d}}(t)=\rho_{\mathrm{E}}^{\mathrm{eq}}(t)$,
and thus $S\left[\rho_{\mathrm{E}}^{\mathrm{d}}(t)||\rho_{\mathrm{E}}^{\mathrm{eq}}(t)\right]=0$.
In this case, Eq. (\ref{eq:Mainresult}) becomes

\begin{equation}
\Delta S_{\mathrm{S}}=\frac{\Delta Q_{\mathrm{S}}}{T_{\mathrm{E}}}+\frac{\Delta Q_{\mathrm{S}}^{2}}{2C_{\mathrm{E}}T_{\mathrm{E}}^{2}}+\Delta\mathscr{C}_{\mathrm{E}}+\Delta I.\label{eq:Mainresult-1}
\end{equation}
Since $\Delta I\geq0$, the explicit form of the second law in quantum
regime can be derived directly from the above equality as 

\begin{equation}
\Delta S_{\mathrm{S}}\geq\frac{\Delta Q_{\mathrm{S}}}{T_{\mathrm{E}}}+\frac{\Delta Q_{\mathrm{S}}^{2}}{2C_{\mathrm{E}}T_{\mathrm{E}}^{2}}+\Delta\mathscr{C}_{\mathrm{E}},\label{eq:Mainresult-1-1}
\end{equation}
where the equal sign is hold if and only if $I(t)=0$, namely, the
final correlation between $\mathrm{S}$ and $\mathrm{E}$ can be ignored
in comparison with other terms. We note that a relevant result was
obtained recently in Ref. \citep{rodrigues2019thermodynamics}, however,
only limited to the case where the reservoirs are described by collisional
models. The result presented here is applicable to the systems coupled
with generic reservoirs.

With the main result of this work illustrated in Eq. (\ref{eq:Mainresult-1-1}),
we summarize the second law of thermodynamics in different regime
in Tab. \ref{tab:The-second-law}. In this table, for the case of
finite reservoir without coherence, similar results were obtained
in some recent studies with specific examples \citep{reeb2015tight,richens2018finite,timpanaro2020landauer}.
In the case that the reservoir is coherent, our result shows that,
in addition to the change in entropy of the system, the change in
coherence of the reservoir also contributes to the amount of heat
exchange between the system and the reservoir. This implies that the
coherence can be used as a quantum resource for energy extraction
in thermodynamic process. Next, we will further show how to take advantages
of this quantum resource to make a heat engine working between two
coherent reservoirs output more work with higher efficiency than its
classical counterpart. 
\begin{table}
\includegraphics[width=8.5cm]{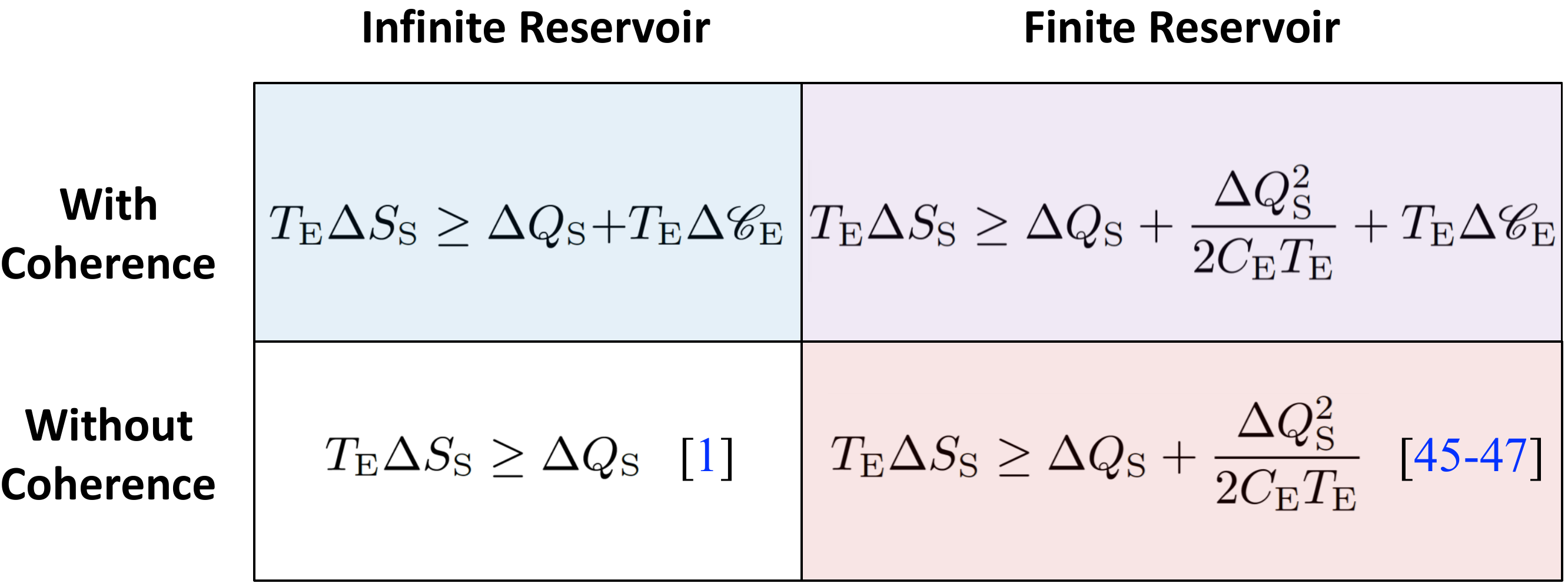}\caption{\label{tab:The-second-law}The second law of thermodynamics in different
regime. Here, in the considered thermodynamic process, $T_{\mathrm{E}}$
and $C_{\mathrm{E}}$ are respectively the effective temperature and
heat capacity of the reservoir. $\Delta S_{\mathrm{S}}$ is the change
in von Neumann entropy of the system, $\Delta Q_{\mathrm{S}}$ is
the heat absorbed from the reservoir, and $\Delta\mathscr{C}_{\mathrm{E}}$
represents the change in coherence of the reservoir with $\mathscr{C}_{\mathrm{E}}$
being the relative entropy of coherence.}
\end{table}

\textit{Surpassing the Carnot efficiency with coherent reservoirs.}-Consider
a general Carnot cycle, consisting of two isothermal and two adiabatic
processes. In the high (low) temperature isothermal process, the working
substance is in contact with the heat reservoir of effective temperature
$T_{\mathrm{h}}$ ($T_{\mathrm{c}}$). Here, we do not limit the specific
characteristics of the working substance. To use Eq. (\ref{eq:Mainresult-1})
to describe the heat change in the isothermal processes, the composite
system of the heat engine and the heat reservoir need to be isolated
from the outside world during the corresponding isothermal process.
In the high temperature and low temperature isothermal process, it
follows from Eq. (\ref{eq:Mainresult-1}) that the heat absorbed and
heat released read $\Delta Q_{\mathrm{h}}=T_{\mathrm{h}}\Delta S_{\mathrm{S}}-T_{\mathrm{h}}\left(\Delta\mathscr{C}_{\mathrm{h}}+\Delta I_{\mathrm{h}}\right)$,
and $\Delta Q_{\mathrm{c}}=-\left[T_{\mathrm{c}}\left(-\Delta S_{\mathrm{S}}\right)-T_{\mathrm{c}}\left(\Delta\mathscr{C}_{\mathrm{c}}+\Delta I_{\mathrm{c}}\right)\right]$,
respectively. Here, we have assumed that the reservoirs are large
enough in comparison with the working substance, thus the finite-size
effect corresponded to the $C_{\mathrm{E}}^{-1}$ term is ignored.
$\Delta S_{\mathrm{S}}>0$ is the entropy change of the working substance
in the high temperature isothermal process, and $\Delta\mathscr{C}_{\mathrm{\alpha}}$
($\alpha=\mathrm{h},\mathrm{c}$) is the change in relative entropy
of coherence of reservoir $\alpha$. To focus on the effect of coherence,
we further assume that the correlation change between the working
substance and the reservoir can be ignored in comparison with the
change in coherence of the reservoir, i.e., $\Delta I\ll\Delta\mathscr{C}$
\citep{DeltaI}. In this sense, according to the definition of the
efficiency $\eta\equiv\left(\Delta Q_{\mathrm{h}}-\Delta Q_{\mathrm{c}}\right)/\Delta Q_{\mathrm{h}}$,
we obtain

\begin{equation}
\eta=\eta_{\mathrm{C}}-\left(1-\eta_{\mathrm{C}}\right)\frac{\Delta\mathscr{C}_{\mathrm{h}}+\Delta\mathscr{C}_{\mathrm{c}}}{\Delta S_{\mathrm{S}}-\Delta\mathscr{C}_{\mathrm{h}}},\label{eq:eta-1}
\end{equation}
where $\eta_{\mathrm{C}}=1-T_{\mathrm{c}}/T_{\mathrm{h}}$ is the
Carnot efficiency, known as the upper bound of efficiency in classical
thermodynamics. Basically, due to the existence of decoherence effect,
the coherence of the reservoirs at the end of the isothermal process
will be smaller than that before the isothermal process \citep{quan2006quantum},
which means $\Delta\mathscr{C}_{\mathrm{h}}<0$ and $\Delta\mathscr{C}_{\mathrm{c}}<0$
(we will illustrate this fact will with a specific example below).
In this case, one has

\begin{equation}
\eta=\eta_{\mathrm{C}}+\left(1-\eta_{\mathrm{C}}\right)\left|\frac{\Delta\mathscr{C}_{\mathrm{h}}+\Delta\mathscr{C}_{\mathrm{c}}}{\Delta S_{\mathrm{S}}-\Delta\mathscr{C}_{\mathrm{h}}}\right|>\eta_{\mathrm{C}},
\end{equation}
which shows the efficiency of the cycle can surpass the Carnot efficiency.
Correspondingly, the work output per cycle $W=W_{c}+\sum_{\alpha=h,c}T_{\mathrm{\alpha}}\left|\Delta\mathscr{C}_{\mathrm{\alpha}}\right|$
is larger than its classical counterpart $W_{c}=\left(T_{\mathrm{h}}-T_{\mathrm{c}}\right)\Delta S_{\mathrm{S}}.$
This means that extra work $W_{e}\equiv W-W_{c}=\sum_{\alpha=h,c}T_{\mathrm{\alpha}}\left|\Delta\mathscr{C}_{\mathrm{\alpha}}\right|$
can be extracted from the reservoirs by utilizing the quantum resource
of coherence. 

Before proceeding further, we give three remarks to the above result:

At first, even there is no effective temperature difference between
the two reservoirs, i.e., $\eta_{\mathrm{C}}=0$, the efficiency of
the cycle $\eta=\left|\Delta\mathscr{C}_{\mathrm{h}}+\Delta\mathscr{C}_{\mathrm{c}}\right|/\left|\Delta S_{\mathrm{S}}-\Delta\mathscr{C}_{\mathrm{h}}\right|>0$
can be non-zero due to the coherence of the reservoirs. This is a
pure quantum effect, because in classical thermodynamics, the heat
engine operating between two heat reservoirs with same temperature
cannot output work. This implies that the quantum coherence alone
can be used to output work. Similar result has been discovered with
a specific model by Scully et al. \citep{scully2003extracting}. However,
their investigation is limited to the cavity-QED system, and their
main finding that the efficiency is higher than the Carnot efficiency
is based on the assumption that the working substance is in the equilibrium
state with a coherence modified temperature \citep{scully2003extracting,quan2006quantum}.
In the current work, we obtain the result from the modified second
law of thermodynamics in quantum regime, without relying on the assumption
that the working substance is in the thermal equilibrium state. Our
results are universal for generic working substances and reservoirs,
and the effect of the reservoirs' coherence is directly demonstrated. 

Secondly, a recent study \citep{francica2019role} connected the coherence
of a driven non-equilibrium quantum system to its irreversible entropy
generation. In this work, we demonstrate that the coherence of the
reservoir, which the system is in contact with, allows the system
to extract more energy from the reservoir to output work. In general,
combining our current work and Ref. \citep{francica2019role} together,
we can conclude that in the thermodynamic of a (interaction) non-equilibrium
quantum system, both of the reservoir's coherence and the system's
coherence are of great significance.\textcolor{black}{{} }

At last, we emphasize here that this result does not violate the second
law thermodynamics in classical regime. After a cycle, although the
state of the working substance returns to its initial state, the states
of the heat reservoirs do not return to their initial states due to
the loss of coherence. The price of the heat engine outputting more
work is the sacrifice of the coherence of the heat reservoirs. If
a larger thermodynamic cycle is considered, namely, the working substance
and the heat reservoirs are all included, we believe that the Carnot
bound would not be violated. Similar idea has been successfully used
to explain the Maxwell's paradox, in which case the Landauer's principle
\citep{landauer1961irreversibility,bennett1982thermodynamics} states
that erasing the memory of the demon requires extra work \citep{dong2011quantum}.
In our case, the preparation of reservoirs' coherence requires additional
energy in principle.

\begin{figure}
\begin{centering}
\includegraphics[width=8.5cm]{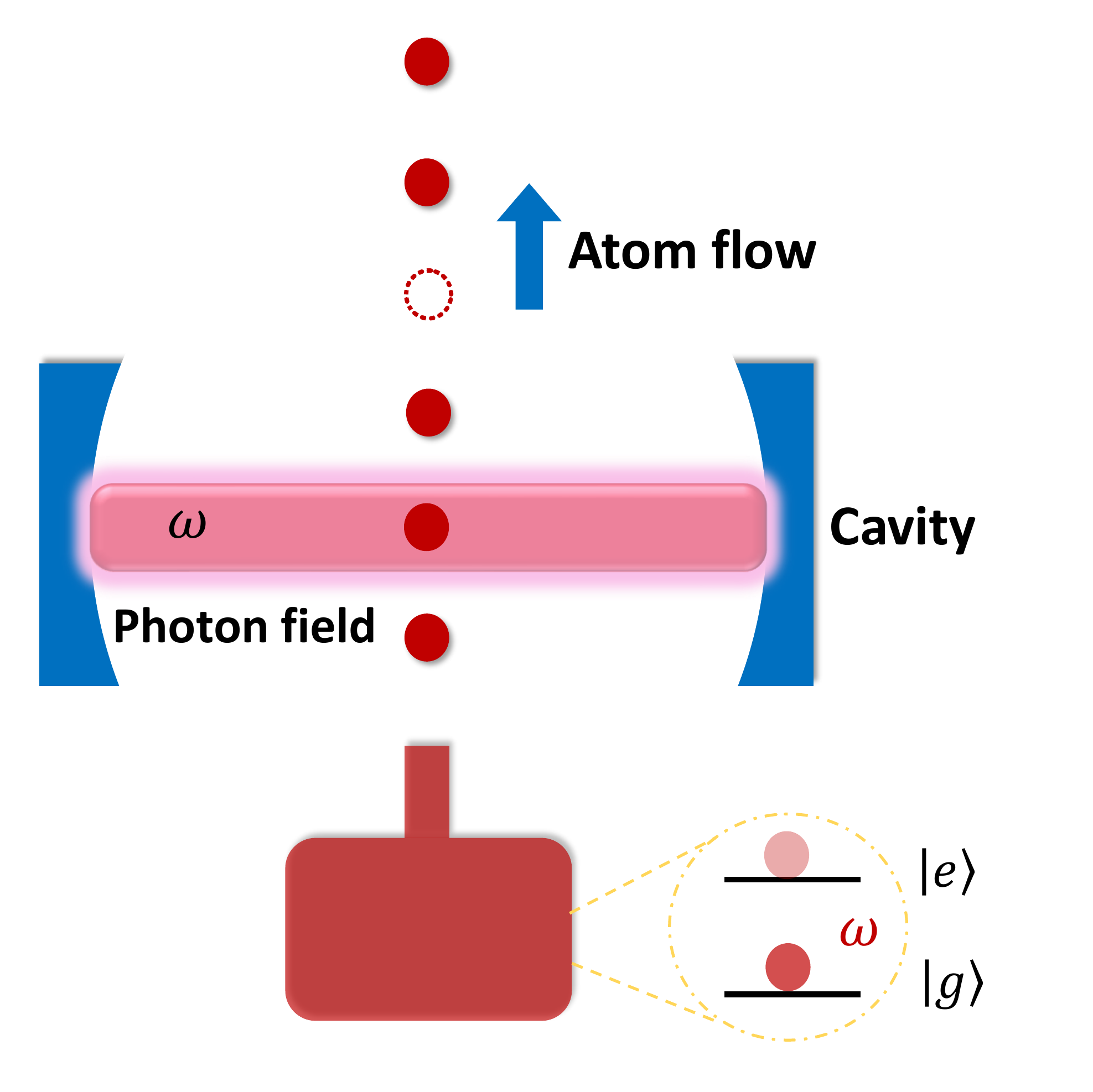}\caption{\label{fig:Schematic-of-the}Schematic of the atoms flow in cavity.
The two-level atoms (energy gap $\omega$), with quantum coherence
in the the excited state $\left|e\right\rangle $ and $\left|g\right\rangle $,
pass through the cavity with rare $r$ and interact with the resonant
photon field in the cavity.}
\par\end{centering}
\end{figure}

\textit{Example.}-We further illustrate the above discussion with
a specific example. Inspired by the studies in \citep{scully2003extracting,quan2006quantum},
we specific our isolated system as a cavity QED system. The whole
system is illustrated in the schematic graph in Fig. \ref{fig:Schematic-of-the}.
As shown in the figure, $N$ atoms pass through the photon field in
the cavity with a rate $r$, and the duration of this thermodynamic
process is $t_{\mathrm{f}}$. We specify the atoms as the reservoir,
and the photon field as the system. We stress here, different from
the model considered in \citep{scully2003extracting,quan2006quantum},
for simplicity, the atoms are assumed to be two-level systems rather
than three-level systems with degeneracy in this work. The atom-photon
interaction is described by the Hamiltonian $V=\hbar g\left|e\right\rangle \left\langle g\right|a+h.c,$
where $\left|e\right\rangle $ ($\left|g\right\rangle $ ) is the
excited (ground) state of the atom, $a$ is the annihilation operator
of the photon field, and $g$ is the coupling constant. The energy
spacing of the two-level atom is $\omega$. For each atom interact
with the photon field in a time interval $\tau=t_{\mathrm{f}}/N=r^{-1}$,
the dynamics of the whole system is dominated by the Hamiltonian of
the Jaynes-Cummings model at resonance \citep{scully1997quantum}.
Suppose each atom is initialized in 
\begin{equation}
\rho_{a}(t)=p_{e}\left|e\right\rangle \left\langle e\right|+p_{g}\left|g\right\rangle \left\langle g\right|+\mu\left|e\right\rangle \left\langle g\right|+\mu^{*}\left|g\right\rangle \left\langle e\right|,
\end{equation}
and the photon field is prepared in a thermal state, thus the evolution
of the whole model can be solved analytically \citep{SM1}. In this
case, the change in coherence of all the $N$ atoms during the isothermal
process $\Delta\xi_{\mathrm{E}}\left(t_{\mathrm{f}}\right)\equiv\xi_{\mathrm{E}}\left(t_{\mathrm{f}}\right)-\xi_{\mathrm{E}}\left(0\right)$
is $\Delta\xi_{\mathrm{E}}(t_{\mathrm{f}})=-\varGamma\xi_{a}(0)t_{\mathrm{f}}$
(See \citep{SM1} for details). Here, $\xi_{a}(0)$ is the initial
relative entropy of coherence of each atom. The coefficient $\varGamma=g^{2}r^{-1}\left(1+2\left\langle n\right\rangle \right)$
is an increasing function of the coupling constant $g$ and the mean
particle number $\left\langle n\right\rangle $ of the photon field,
and a decreasing function of the passing rate $r$ of the atoms. Then,
according to Eq. (\ref{eq:eta-1}), the efficiency of this cycle is
explicitly obtained as \citep{SM1}

\begin{equation}
\eta=\eta_{\mathrm{C}}+\left(1-\eta_{\mathrm{C}}\right)\frac{\varGamma\xi_{h}t_{h}}{\Delta S_{l}+\varGamma\xi_{h}t_{h}}.\label{eq:eta1}
\end{equation}
Here, we have assumed that in the high-temperature isothermal process
of duration $t_{h}$, the relative entropy of coherence of each initial
atom is $\xi_{a}(0)=\xi_{h}$, while the atoms in the low-temperature
isothermal process have vanish coherence, namely, $\xi_{c}=0$. $\Delta S_{l}$
is the change in entropy of the photon field per cycle. Obviously,
Eq. (\ref{eq:eta1}) indicates that the efficiency of such photon-cavity
Carnot engine can surpass the Carnot efficiency, which is defined
by the effective temperature of the atoms in the two isothermal processes.
And in the long-time limit of $\varGamma\xi_{h}t_{h}\gg\Delta S_{l}$,
the efficiency can approach $1$, independent of the corresponding
Carnot efficiency. Besides, note that for a given $t_{\mathrm{f}}$,
with the increasing of atom injection rate $r$, the coherence correction
term in Eq. (\ref{eq:eta1}) decreases until it reaches zero.

\textit{Conclusions and discussions.}\textbf{\textit{-}}As a summary,
we studied the second law of thermodynamics in quantum regime, which
involves a system interacting with a finite-sized reservoir with quantum
coherence characterized by off-diagonal terms in the density matrix.
It is shown that the coherence, qualified by relative entropy, can
serve as a useful resource for improving the output efficiency of
heat engines. Particularly, even there is no temperature difference
between the two reservoirs, the heat engine can also output non-vanishing
work resulting from the reservoir's coherence. The above observations
were further demonstrated with a photon-cavity-Carnot engine, where
the reservoir consists of two-level atoms with coherence\textcolor{black}{.
It should be mentioned that, besides the relative entropy of coherence
\citep{Quantifying2014}, other coherence measures had also been proposed,
such as the $l_{1}$ norm of coherence \citep{Quantifying2014}, the
robustness of coherence \citep{napoli2016robustness}, and the coherence
of formation \citep{yuan2015intrinsic}. Nevertheless, our current
studies revealed that only the relative entropy of coherence is feasible
to describe energy exchange in the thermodynamics with quantum coherence.
The coherence-enhancement in energy conversion processes predicted
above can be tested in some state-of-art experiment}s for quantum
thermodynamics \citep{Rossnagel2016,shortcutHEexperimentSciAdv2018,atomic-Endoreversible2020}.

The general findings in the current study are expected to be applied
to optimize the operation of quantum heat engine in finite-time cycles,
although the optimizations for classical heat engine have been extensively
investigated \citep{Holubec2017,kosloff2019quantum,tu2021abstract}.
We suggest to consider the relevant issues listed as follows: (i)
the influence of the reservoir coherence on the engine efficiency
at maximum power \citep{chen2019boosting}, the power-efficiency constraint
relation \citep{Constraintrelationyhma}, and the efficiency fluctuation
\citep{fei2021efficiency}; (ii) whether the coherence will change
the optimal operation scheme of the heat engine \citep{yhmaoptimalcontrol,2020Optimal,2020IEGyhma,ma2020self};
(iii) the information correlation \citep{zhang2009hidden,ma2018non}
result from the finite-sized reservoir with non-canonical statistics
\citep{parikh2000hawking,xu2014noncanonical,richens2018finite,ma2018non,ma2018dark}.
The above mentioned investigations in future would deepen the understanding
of the information-assisted thermodynamic processes of energy conservation
in micro-scale.

\textit{Acknowledgment.-}This work is supported by the National Natural
Science Foundation of China (NSFC) (Grants No. 11534002, No. 11875049,
No. U1730449, No. U1530401, and No. U1930403), the National Basic
Research Program of China (Grants No. 2016YFA0301201), and the China
Postdoctoral Science Foundation (Grant No. BX2021030).

\bibliographystyle{apsrev}
\addcontentsline{toc}{section}{\refname}\bibliography{QRCT}

\begin{thebibliography}{71}
\expandafter\ifx\csname natexlab\endcsname\relax\def\natexlab#1{#1}\fi
\expandafter\ifx\csname bibnamefont\endcsname\relax
  \def\bibnamefont#1{#1}\fi
\expandafter\ifx\csname bibfnamefont\endcsname\relax
  \def\bibfnamefont#1{#1}\fi
\expandafter\ifx\csname citenamefont\endcsname\relax
  \def\citenamefont#1{#1}\fi
\expandafter\ifx\csname url\endcsname\relax
  \def\url#1{\texttt{#1}}\fi
\expandafter\ifx\csname urlprefix\endcsname\relax\def\urlprefix{URL }\fi
\providecommand{\bibinfo}[2]{#2}
\providecommand{\eprint}[2][]{\url{#2}}

\bibitem[{\citenamefont{Huang}(2013)}]{Carnoteff}
\bibinfo{author}{\bibfnamefont{K.}~\bibnamefont{Huang}},
  \emph{\bibinfo{title}{Introduction To Statistical Physics, 2Nd Edition}}
  (\bibinfo{publisher}{T\&F/Crc Press}, \bibinfo{year}{2013}), ISBN
  \bibinfo{isbn}{978-1-4200-7902-9}.

\bibitem[{\citenamefont{Esposito et~al.}(2009)\citenamefont{Esposito, Harbola,
  and Mukamel}}]{Esposito_2009}
\bibinfo{author}{\bibfnamefont{M.}~\bibnamefont{Esposito}},
  \bibinfo{author}{\bibfnamefont{U.}~\bibnamefont{Harbola}}, \bibnamefont{and}
  \bibinfo{author}{\bibfnamefont{S.}~\bibnamefont{Mukamel}},
  \bibinfo{journal}{Rev. Mod. Phys.} \textbf{\bibinfo{volume}{81}},
  \bibinfo{pages}{1665} (\bibinfo{year}{2009}).

\bibitem[{\citenamefont{Campisi et~al.}(2011)\citenamefont{Campisi, H\"{a}nggi,
  and Talkner}}]{Campisi_2011}
\bibinfo{author}{\bibfnamefont{M.}~\bibnamefont{Campisi}},
  \bibinfo{author}{\bibfnamefont{P.}~\bibnamefont{H\"{a}nggi}},
  \bibnamefont{and} \bibinfo{author}{\bibfnamefont{P.}~\bibnamefont{Talkner}},
  \bibinfo{journal}{Rev. Mod. Phys.} \textbf{\bibinfo{volume}{83}},
  \bibinfo{pages}{771} (\bibinfo{year}{2011}).

\bibitem[{\citenamefont{Goold et~al.}(2016)\citenamefont{Goold, Huber, Riera,
  Del~Rio, and Skrzypczyk}}]{goold2016role}
\bibinfo{author}{\bibfnamefont{J.}~\bibnamefont{Goold}},
  \bibinfo{author}{\bibfnamefont{M.}~\bibnamefont{Huber}},
  \bibinfo{author}{\bibfnamefont{A.}~\bibnamefont{Riera}},
  \bibinfo{author}{\bibfnamefont{L.}~\bibnamefont{Del~Rio}}, \bibnamefont{and}
  \bibinfo{author}{\bibfnamefont{P.}~\bibnamefont{Skrzypczyk}},
  \bibinfo{journal}{J. Phys. A: Math. and Theor.}
  \textbf{\bibinfo{volume}{49}}, \bibinfo{pages}{143001}
  (\bibinfo{year}{2016}).

\bibitem[{\citenamefont{Streltsov et~al.}(2017)\citenamefont{Streltsov, Adesso,
  and Plenio}}]{streltsov2017colloquium}
\bibinfo{author}{\bibfnamefont{A.}~\bibnamefont{Streltsov}},
  \bibinfo{author}{\bibfnamefont{G.}~\bibnamefont{Adesso}}, \bibnamefont{and}
  \bibinfo{author}{\bibfnamefont{M.~B.} \bibnamefont{Plenio}},
  \bibinfo{journal}{Rev. Mod. Phys.} \textbf{\bibinfo{volume}{89}},
  \bibinfo{pages}{041003} (\bibinfo{year}{2017}).

\bibitem[{\citenamefont{Vinjanampathy and Anders}(2016)}]{Vinjanampathy_2016}
\bibinfo{author}{\bibfnamefont{S.}~\bibnamefont{Vinjanampathy}}
  \bibnamefont{and} \bibinfo{author}{\bibfnamefont{J.}~\bibnamefont{Anders}},
  \bibinfo{journal}{Contemp. Phys.} \textbf{\bibinfo{volume}{57}},
  \bibinfo{pages}{545} (\bibinfo{year}{2016}).

\bibitem[{\citenamefont{Binder et~al.}(2018)\citenamefont{Binder, Correa,
  Gogolin, Anders, and Adesso}}]{Binder2018}
\bibinfo{editor}{\bibfnamefont{F.}~\bibnamefont{Binder}},
  \bibinfo{editor}{\bibfnamefont{L.~A.} \bibnamefont{Correa}},
  \bibinfo{editor}{\bibfnamefont{C.}~\bibnamefont{Gogolin}},
  \bibinfo{editor}{\bibfnamefont{J.}~\bibnamefont{Anders}}, \bibnamefont{and}
  \bibinfo{editor}{\bibfnamefont{G.}~\bibnamefont{Adesso}}, eds.,
  \emph{\bibinfo{title}{Thermodynamics in the Quantum Regime}}
  (\bibinfo{publisher}{Springer International Publishing},
  \bibinfo{year}{2018}).

\bibitem[{\citenamefont{Kosloff}(2019)}]{kosloff2019quantum}
\bibinfo{author}{\bibfnamefont{R.}~\bibnamefont{Kosloff}}, \bibinfo{journal}{J.
  Chem. Phys.} \textbf{\bibinfo{volume}{150}}, \bibinfo{pages}{204105}
  (\bibinfo{year}{2019}).

\bibitem[{\citenamefont{Su et~al.}(2021{\natexlab{a}})\citenamefont{Su, Zhang,
  Peng, Su, and Chen}}]{su2021thermodynamic}
\bibinfo{author}{\bibfnamefont{S.}~\bibnamefont{Su}},
  \bibinfo{author}{\bibfnamefont{Y.}~\bibnamefont{Zhang}},
  \bibinfo{author}{\bibfnamefont{W.}~\bibnamefont{Peng}},
  \bibinfo{author}{\bibfnamefont{G.}~\bibnamefont{Su}}, \bibnamefont{and}
  \bibinfo{author}{\bibfnamefont{J.}~\bibnamefont{Chen}},
  \bibinfo{journal}{SCIENTIA SINICA Physica, Mechanica \& Astronomica}
  \textbf{\bibinfo{volume}{51}}, \bibinfo{pages}{030011}
  (\bibinfo{year}{2021}{\natexlab{a}}).

\bibitem[{\citenamefont{Brandao et~al.}(2013)\citenamefont{Brandao, Horodecki,
  Oppenheim, Renes, and Spekkens}}]{brandao2013resource}
\bibinfo{author}{\bibfnamefont{F.~G.} \bibnamefont{Brandao}},
  \bibinfo{author}{\bibfnamefont{M.}~\bibnamefont{Horodecki}},
  \bibinfo{author}{\bibfnamefont{J.}~\bibnamefont{Oppenheim}},
  \bibinfo{author}{\bibfnamefont{J.~M.} \bibnamefont{Renes}}, \bibnamefont{and}
  \bibinfo{author}{\bibfnamefont{R.~W.} \bibnamefont{Spekkens}},
  \bibinfo{journal}{Phys. Rev. Lett.} \textbf{\bibinfo{volume}{111}},
  \bibinfo{pages}{250404} (\bibinfo{year}{2013}).

\bibitem[{\citenamefont{Gour et~al.}(2015)\citenamefont{Gour, M{\"u}ller,
  Narasimhachar, Spekkens, and Halpern}}]{gour2015resource}
\bibinfo{author}{\bibfnamefont{G.}~\bibnamefont{Gour}},
  \bibinfo{author}{\bibfnamefont{M.~P.} \bibnamefont{M{\"u}ller}},
  \bibinfo{author}{\bibfnamefont{V.}~\bibnamefont{Narasimhachar}},
  \bibinfo{author}{\bibfnamefont{R.~W.} \bibnamefont{Spekkens}},
  \bibnamefont{and} \bibinfo{author}{\bibfnamefont{N.~Y.}
  \bibnamefont{Halpern}}, \bibinfo{journal}{Phys. Rep.}
  \textbf{\bibinfo{volume}{583}}, \bibinfo{pages}{1} (\bibinfo{year}{2015}).

\bibitem[{\citenamefont{Sparaciari et~al.}(2017)\citenamefont{Sparaciari,
  Oppenheim, and Fritz}}]{sparaciari2017resource}
\bibinfo{author}{\bibfnamefont{C.}~\bibnamefont{Sparaciari}},
  \bibinfo{author}{\bibfnamefont{J.}~\bibnamefont{Oppenheim}},
  \bibnamefont{and} \bibinfo{author}{\bibfnamefont{T.}~\bibnamefont{Fritz}},
  \bibinfo{journal}{Phys. Rev. A} \textbf{\bibinfo{volume}{96}},
  \bibinfo{pages}{052112} (\bibinfo{year}{2017}).

\bibitem[{\citenamefont{Lostaglio}(2019)}]{lostaglio2019introductory}
\bibinfo{author}{\bibfnamefont{M.}~\bibnamefont{Lostaglio}},
  \bibinfo{journal}{Rep. Prog. Phys.} \textbf{\bibinfo{volume}{82}},
  \bibinfo{pages}{114001} (\bibinfo{year}{2019}).

\bibitem[{\citenamefont{Alicki and Fannes}(2013)}]{alicki2013entanglement}
\bibinfo{author}{\bibfnamefont{R.}~\bibnamefont{Alicki}} \bibnamefont{and}
  \bibinfo{author}{\bibfnamefont{M.}~\bibnamefont{Fannes}},
  \bibinfo{journal}{Phys. Rev. E} \textbf{\bibinfo{volume}{87}},
  \bibinfo{pages}{042123} (\bibinfo{year}{2013}).

\bibitem[{\citenamefont{Campisi and Fazio}(2016)}]{campisi2016power}
\bibinfo{author}{\bibfnamefont{M.}~\bibnamefont{Campisi}} \bibnamefont{and}
  \bibinfo{author}{\bibfnamefont{R.}~\bibnamefont{Fazio}},
  \bibinfo{journal}{Nat. Comm.} \textbf{\bibinfo{volume}{7}},
  \bibinfo{pages}{11895} (\bibinfo{year}{2016}).

\bibitem[{\citenamefont{Ma et~al.}(2017)\citenamefont{Ma, Su, and
  Sun}}]{YHMaQPTHE}
\bibinfo{author}{\bibfnamefont{Y.-H.} \bibnamefont{Ma}},
  \bibinfo{author}{\bibfnamefont{S.-H.} \bibnamefont{Su}}, \bibnamefont{and}
  \bibinfo{author}{\bibfnamefont{C.-P.} \bibnamefont{Sun}},
  \bibinfo{journal}{Phys. Rev. E} \textbf{\bibinfo{volume}{96}},
  \bibinfo{pages}{022143} (\bibinfo{year}{2017}).

\bibitem[{\citenamefont{Bresque et~al.}(2021)\citenamefont{Bresque, Camati,
  Rogers, Murch, Jordan, and Auff{\`e}ves}}]{bresque2021two}
\bibinfo{author}{\bibfnamefont{L.}~\bibnamefont{Bresque}},
  \bibinfo{author}{\bibfnamefont{P.~A.} \bibnamefont{Camati}},
  \bibinfo{author}{\bibfnamefont{S.}~\bibnamefont{Rogers}},
  \bibinfo{author}{\bibfnamefont{K.}~\bibnamefont{Murch}},
  \bibinfo{author}{\bibfnamefont{A.~N.} \bibnamefont{Jordan}},
  \bibnamefont{and}
  \bibinfo{author}{\bibfnamefont{A.}~\bibnamefont{Auff{\`e}ves}},
  \bibinfo{journal}{Phys. Rev. Lett.} \textbf{\bibinfo{volume}{126}},
  \bibinfo{pages}{120605} (\bibinfo{year}{2021}).

\bibitem[{\citenamefont{Scully et~al.}(2003)\citenamefont{Scully, Zubairy,
  Agarwal, and Walther}}]{scully2003extracting}
\bibinfo{author}{\bibfnamefont{M.~O.} \bibnamefont{Scully}},
  \bibinfo{author}{\bibfnamefont{M.~S.} \bibnamefont{Zubairy}},
  \bibinfo{author}{\bibfnamefont{G.~S.} \bibnamefont{Agarwal}},
  \bibnamefont{and} \bibinfo{author}{\bibfnamefont{H.}~\bibnamefont{Walther}},
  \bibinfo{journal}{Science} \textbf{\bibinfo{volume}{299}},
  \bibinfo{pages}{862} (\bibinfo{year}{2003}).

\bibitem[{\citenamefont{Quan et~al.}(2006)\citenamefont{Quan, Zhang, and
  Sun}}]{quan2006quantum}
\bibinfo{author}{\bibfnamefont{H.}~\bibnamefont{Quan}},
  \bibinfo{author}{\bibfnamefont{P.}~\bibnamefont{Zhang}}, \bibnamefont{and}
  \bibinfo{author}{\bibfnamefont{C.}~\bibnamefont{Sun}},
  \bibinfo{journal}{Phys. Rev. E} \textbf{\bibinfo{volume}{73}},
  \bibinfo{pages}{036122} (\bibinfo{year}{2006}).

\bibitem[{\citenamefont{Scully et~al.}(2011)\citenamefont{Scully, Chapin,
  Dorfman, Kim, and Svidzinsky}}]{scully2011quantum}
\bibinfo{author}{\bibfnamefont{M.~O.} \bibnamefont{Scully}},
  \bibinfo{author}{\bibfnamefont{K.~R.} \bibnamefont{Chapin}},
  \bibinfo{author}{\bibfnamefont{K.~E.} \bibnamefont{Dorfman}},
  \bibinfo{author}{\bibfnamefont{M.~B.} \bibnamefont{Kim}}, \bibnamefont{and}
  \bibinfo{author}{\bibfnamefont{A.}~\bibnamefont{Svidzinsky}},
  \bibinfo{journal}{Proc. Natl. Acad. Sci.} \textbf{\bibinfo{volume}{108}},
  \bibinfo{pages}{15097} (\bibinfo{year}{2011}).

\bibitem[{\citenamefont{Park et~al.}(2013)\citenamefont{Park, Kim, Sagawa, and
  Kim}}]{park2013heat}
\bibinfo{author}{\bibfnamefont{J.~J.} \bibnamefont{Park}},
  \bibinfo{author}{\bibfnamefont{K.-H.} \bibnamefont{Kim}},
  \bibinfo{author}{\bibfnamefont{T.}~\bibnamefont{Sagawa}}, \bibnamefont{and}
  \bibinfo{author}{\bibfnamefont{S.~W.} \bibnamefont{Kim}},
  \bibinfo{journal}{Phys. Rev. Lett.} \textbf{\bibinfo{volume}{111}},
  \bibinfo{pages}{230402} (\bibinfo{year}{2013}).

\bibitem[{\citenamefont{Ro{\ss}nagel et~al.}(2014)\citenamefont{Ro{\ss}nagel,
  Abah, Schmidt-Kaler, Singer, and Lutz}}]{NanoscaleHEPRL2014}
\bibinfo{author}{\bibfnamefont{J.}~\bibnamefont{Ro{\ss}nagel}},
  \bibinfo{author}{\bibfnamefont{O.}~\bibnamefont{Abah}},
  \bibinfo{author}{\bibfnamefont{F.}~\bibnamefont{Schmidt-Kaler}},
  \bibinfo{author}{\bibfnamefont{K.}~\bibnamefont{Singer}}, \bibnamefont{and}
  \bibinfo{author}{\bibfnamefont{E.}~\bibnamefont{Lutz}},
  \bibinfo{journal}{Phys. Rev. Lett.} \textbf{\bibinfo{volume}{112}}
  (\bibinfo{year}{2014}).

\bibitem[{\citenamefont{Uzdin et~al.}(2015)\citenamefont{Uzdin, Levy, and
  Kosloff}}]{uzdin2015equivalence}
\bibinfo{author}{\bibfnamefont{R.}~\bibnamefont{Uzdin}},
  \bibinfo{author}{\bibfnamefont{A.}~\bibnamefont{Levy}}, \bibnamefont{and}
  \bibinfo{author}{\bibfnamefont{R.}~\bibnamefont{Kosloff}},
  \bibinfo{journal}{Phys. Rev. X} \textbf{\bibinfo{volume}{5}},
  \bibinfo{pages}{031044} (\bibinfo{year}{2015}).

\bibitem[{\citenamefont{Dorfman et~al.}(2018)\citenamefont{Dorfman, Xu, and
  Cao}}]{dorfman2018efficiency}
\bibinfo{author}{\bibfnamefont{K.~E.} \bibnamefont{Dorfman}},
  \bibinfo{author}{\bibfnamefont{D.}~\bibnamefont{Xu}}, \bibnamefont{and}
  \bibinfo{author}{\bibfnamefont{J.}~\bibnamefont{Cao}},
  \bibinfo{journal}{Phys. Rev. E} \textbf{\bibinfo{volume}{97}},
  \bibinfo{pages}{042120} (\bibinfo{year}{2018}).

\bibitem[{\citenamefont{Camati et~al.}(2019)\citenamefont{Camati, Santos, and
  Serra}}]{Camati2019}
\bibinfo{author}{\bibfnamefont{P.~A.} \bibnamefont{Camati}},
  \bibinfo{author}{\bibfnamefont{J.~F.~G.} \bibnamefont{Santos}},
  \bibnamefont{and} \bibinfo{author}{\bibfnamefont{R.~M.} \bibnamefont{Serra}},
  \bibinfo{journal}{Phys. Rev. A} \textbf{\bibinfo{volume}{99}}
  (\bibinfo{year}{2019}).

\bibitem[{\citenamefont{Su et~al.}(2021{\natexlab{b}})\citenamefont{Su, Zhang,
  Andresen, and Chen}}]{su2021coherence}
\bibinfo{author}{\bibfnamefont{S.}~\bibnamefont{Su}},
  \bibinfo{author}{\bibfnamefont{Y.}~\bibnamefont{Zhang}},
  \bibinfo{author}{\bibfnamefont{B.}~\bibnamefont{Andresen}}, \bibnamefont{and}
  \bibinfo{author}{\bibfnamefont{J.}~\bibnamefont{Chen}},
  \bibinfo{journal}{Physica A: Statistical Mechanics and its Applications}
  \textbf{\bibinfo{volume}{569}}, \bibinfo{pages}{125753}
  (\bibinfo{year}{2021}{\natexlab{b}}).

\bibitem[{\citenamefont{Lostaglio et~al.}(2015)\citenamefont{Lostaglio,
  Jennings, and Rudolph}}]{lostaglio2015description}
\bibinfo{author}{\bibfnamefont{M.}~\bibnamefont{Lostaglio}},
  \bibinfo{author}{\bibfnamefont{D.}~\bibnamefont{Jennings}}, \bibnamefont{and}
  \bibinfo{author}{\bibfnamefont{T.}~\bibnamefont{Rudolph}},
  \bibinfo{journal}{Nat. Comm.} \textbf{\bibinfo{volume}{6}},
  \bibinfo{pages}{1} (\bibinfo{year}{2015}).

\bibitem[{\citenamefont{Kammerlander and
  Anders}(2016)}]{kammerlander2016coherence}
\bibinfo{author}{\bibfnamefont{P.}~\bibnamefont{Kammerlander}}
  \bibnamefont{and} \bibinfo{author}{\bibfnamefont{J.}~\bibnamefont{Anders}},
  \bibinfo{journal}{Sci. Rep.} \textbf{\bibinfo{volume}{6}}, \bibinfo{pages}{1}
  (\bibinfo{year}{2016}).

\bibitem[{\citenamefont{Korzekwa et~al.}(2016)\citenamefont{Korzekwa,
  Lostaglio, Oppenheim, and Jennings}}]{korzekwa2016extraction}
\bibinfo{author}{\bibfnamefont{K.}~\bibnamefont{Korzekwa}},
  \bibinfo{author}{\bibfnamefont{M.}~\bibnamefont{Lostaglio}},
  \bibinfo{author}{\bibfnamefont{J.}~\bibnamefont{Oppenheim}},
  \bibnamefont{and} \bibinfo{author}{\bibfnamefont{D.}~\bibnamefont{Jennings}},
  \bibinfo{journal}{New J. Phys.} \textbf{\bibinfo{volume}{18}},
  \bibinfo{pages}{023045} (\bibinfo{year}{2016}).

\bibitem[{\citenamefont{Su et~al.}(2018)\citenamefont{Su, Chen, Ma, Chen, and
  Sun}}]{Su2018}
\bibinfo{author}{\bibfnamefont{S.}~\bibnamefont{Su}},
  \bibinfo{author}{\bibfnamefont{J.}~\bibnamefont{Chen}},
  \bibinfo{author}{\bibfnamefont{Y.}~\bibnamefont{Ma}},
  \bibinfo{author}{\bibfnamefont{J.}~\bibnamefont{Chen}}, \bibnamefont{and}
  \bibinfo{author}{\bibfnamefont{C.}~\bibnamefont{Sun}},
  \bibinfo{journal}{Chin. Phys. B} \textbf{\bibinfo{volume}{27}},
  \bibinfo{pages}{060502} (\bibinfo{year}{2018}).

\bibitem[{\citenamefont{Francica et~al.}(2019)\citenamefont{Francica, Goold,
  and Plastina}}]{francica2019role}
\bibinfo{author}{\bibfnamefont{G.}~\bibnamefont{Francica}},
  \bibinfo{author}{\bibfnamefont{J.}~\bibnamefont{Goold}}, \bibnamefont{and}
  \bibinfo{author}{\bibfnamefont{F.}~\bibnamefont{Plastina}},
  \bibinfo{journal}{Phys. Rev. E} \textbf{\bibinfo{volume}{99}},
  \bibinfo{pages}{042105} (\bibinfo{year}{2019}).

\bibitem[{\citenamefont{Rodrigues et~al.}(2019)\citenamefont{Rodrigues,
  De~Chiara, Paternostro, and Landi}}]{rodrigues2019thermodynamics}
\bibinfo{author}{\bibfnamefont{F.~L.} \bibnamefont{Rodrigues}},
  \bibinfo{author}{\bibfnamefont{G.}~\bibnamefont{De~Chiara}},
  \bibinfo{author}{\bibfnamefont{M.}~\bibnamefont{Paternostro}},
  \bibnamefont{and} \bibinfo{author}{\bibfnamefont{G.~T.} \bibnamefont{Landi}},
  \bibinfo{journal}{Phys. Rev. Lett.} \textbf{\bibinfo{volume}{123}},
  \bibinfo{pages}{140601} (\bibinfo{year}{2019}).

\bibitem[{\citenamefont{de~Lima~Bernardo}(2020)}]{de2020unraveling}
\bibinfo{author}{\bibfnamefont{B.}~\bibnamefont{de~Lima~Bernardo}},
  \bibinfo{journal}{Phys. Rev. E} \textbf{\bibinfo{volume}{102}},
  \bibinfo{pages}{062152} (\bibinfo{year}{2020}).

\bibitem[{\citenamefont{Baumgratz et~al.}(2014)\citenamefont{Baumgratz, Cramer,
  and Plenio}}]{Quantifying2014}
\bibinfo{author}{\bibfnamefont{T.}~\bibnamefont{Baumgratz}},
  \bibinfo{author}{\bibfnamefont{M.}~\bibnamefont{Cramer}}, \bibnamefont{and}
  \bibinfo{author}{\bibfnamefont{M.}~\bibnamefont{Plenio}},
  \bibinfo{journal}{Phys. Rev. Lett.} \textbf{\bibinfo{volume}{113}},
  \bibinfo{pages}{140401} (\bibinfo{year}{2014}).

\bibitem[{\citenamefont{Rivas}(2020)}]{rivas2020strong}
\bibinfo{author}{\bibfnamefont{{\'A}.}~\bibnamefont{Rivas}},
  \bibinfo{journal}{Phys. Rev. Lett.} \textbf{\bibinfo{volume}{124}},
  \bibinfo{pages}{160601} (\bibinfo{year}{2020}).

\bibitem[{\citenamefont{Nielsen and Chuang}(2002)}]{nielsen2002quantum}
\bibinfo{author}{\bibfnamefont{M.~A.} \bibnamefont{Nielsen}} \bibnamefont{and}
  \bibinfo{author}{\bibfnamefont{I.}~\bibnamefont{Chuang}},
  \emph{\bibinfo{title}{Quantum computation and quantum information}}
  (\bibinfo{year}{2002}).

\bibitem[{\citenamefont{Li et~al.}(2017)}]{li2017production}
\bibinfo{author}{\bibfnamefont{S.-W.} \bibnamefont{Li}} \bibnamefont{et~al.},
  \bibinfo{journal}{Physical Review E} \textbf{\bibinfo{volume}{96}},
  \bibinfo{pages}{012139} (\bibinfo{year}{2017}).

\bibitem[{SM1()}]{SM1}
\bibinfo{note}{See Supplemental Material for the derivation of Eqs. (7) and
  (10), and the detailed solution of the dynamic ecolution of the
  Jaynes-Cummings model.}

\bibitem[{\citenamefont{Spohn}(1978)}]{spohn1978entropy}
\bibinfo{author}{\bibfnamefont{H.}~\bibnamefont{Spohn}},
  \bibinfo{journal}{Journal of Mathematical Physics}
  \textbf{\bibinfo{volume}{19}}, \bibinfo{pages}{1227} (\bibinfo{year}{1978}).

\bibitem[{\citenamefont{Ma et~al.}(2018{\natexlab{a}})\citenamefont{Ma, Xu,
  Dong, and Sun}}]{Constraintrelationyhma}
\bibinfo{author}{\bibfnamefont{Y.-H.} \bibnamefont{Ma}},
  \bibinfo{author}{\bibfnamefont{D.}~\bibnamefont{Xu}},
  \bibinfo{author}{\bibfnamefont{H.}~\bibnamefont{Dong}}, \bibnamefont{and}
  \bibinfo{author}{\bibfnamefont{C.-P.} \bibnamefont{Sun}},
  \bibinfo{journal}{Phys. Rev. E} \textbf{\bibinfo{volume}{98}},
  \bibinfo{pages}{042112} (\bibinfo{year}{2018}{\natexlab{a}}).

\bibitem[{\citenamefont{Ma et~al.}(2020)\citenamefont{Ma, Zhai, Chen, Dong, and
  Sun}}]{2020IEGyhma}
\bibinfo{author}{\bibfnamefont{Y.-H.} \bibnamefont{Ma}},
  \bibinfo{author}{\bibfnamefont{R.-X.} \bibnamefont{Zhai}},
  \bibinfo{author}{\bibfnamefont{J.}~\bibnamefont{Chen}},
  \bibinfo{author}{\bibfnamefont{H.}~\bibnamefont{Dong}}, \bibnamefont{and}
  \bibinfo{author}{\bibfnamefont{C.~P.} \bibnamefont{Sun}},
  \bibinfo{journal}{Phys. Rev. Lett.} \textbf{\bibinfo{volume}{125}},
  \bibinfo{pages}{210601} (\bibinfo{year}{2020}).

\bibitem[{\citenamefont{Horodecki and
  Oppenheim}(2013)}]{horodecki2013quantumness}
\bibinfo{author}{\bibfnamefont{M.}~\bibnamefont{Horodecki}} \bibnamefont{and}
  \bibinfo{author}{\bibfnamefont{J.}~\bibnamefont{Oppenheim}},
  \bibinfo{journal}{Int. J. Mod. Phys. B} \textbf{\bibinfo{volume}{27}},
  \bibinfo{pages}{1345019} (\bibinfo{year}{2013}).

\bibitem[{\citenamefont{Brandao and Gour}(2015)}]{brandao2015reversible}
\bibinfo{author}{\bibfnamefont{F.~G.} \bibnamefont{Brandao}} \bibnamefont{and}
  \bibinfo{author}{\bibfnamefont{G.}~\bibnamefont{Gour}},
  \bibinfo{journal}{Phys. Rev. Lett.} \textbf{\bibinfo{volume}{115}},
  \bibinfo{pages}{070503} (\bibinfo{year}{2015}).

\bibitem[{\citenamefont{Chitambar and Gour}(2019)}]{chitambar2019quantum}
\bibinfo{author}{\bibfnamefont{E.}~\bibnamefont{Chitambar}} \bibnamefont{and}
  \bibinfo{author}{\bibfnamefont{G.}~\bibnamefont{Gour}},
  \bibinfo{journal}{Rev. Mod. Phys.} \textbf{\bibinfo{volume}{91}},
  \bibinfo{pages}{025001} (\bibinfo{year}{2019}).

\bibitem[{\citenamefont{Reeb and Wolf}(2015)}]{reeb2015tight}
\bibinfo{author}{\bibfnamefont{D.}~\bibnamefont{Reeb}} \bibnamefont{and}
  \bibinfo{author}{\bibfnamefont{M.~M.} \bibnamefont{Wolf}},
  \bibinfo{journal}{IEEE Transactions on Information Theory}
  \textbf{\bibinfo{volume}{61}}, \bibinfo{pages}{1458} (\bibinfo{year}{2015}).

\bibitem[{\citenamefont{Richens et~al.}(2018)\citenamefont{Richens, Alhambra,
  and Masanes}}]{richens2018finite}
\bibinfo{author}{\bibfnamefont{J.~G.} \bibnamefont{Richens}},
  \bibinfo{author}{\bibfnamefont{{\'A}.~M.} \bibnamefont{Alhambra}},
  \bibnamefont{and} \bibinfo{author}{\bibfnamefont{L.}~\bibnamefont{Masanes}},
  \bibinfo{journal}{Phys. Rev. E} \textbf{\bibinfo{volume}{97}},
  \bibinfo{pages}{062132} (\bibinfo{year}{2018}).

\bibitem[{\citenamefont{Timpanaro et~al.}(2020)\citenamefont{Timpanaro, Santos,
  and Landi}}]{timpanaro2020landauer}
\bibinfo{author}{\bibfnamefont{A.~M.} \bibnamefont{Timpanaro}},
  \bibinfo{author}{\bibfnamefont{J.~P.} \bibnamefont{Santos}},
  \bibnamefont{and} \bibinfo{author}{\bibfnamefont{G.~T.} \bibnamefont{Landi}},
  \bibinfo{journal}{Phys. Rev. Lett.} \textbf{\bibinfo{volume}{124}},
  \bibinfo{pages}{240601} (\bibinfo{year}{2020}).

\bibitem[{\citenamefont{Izumida and Okuda}(2014)}]{izumida2014work}
\bibinfo{author}{\bibfnamefont{Y.}~\bibnamefont{Izumida}} \bibnamefont{and}
  \bibinfo{author}{\bibfnamefont{K.}~\bibnamefont{Okuda}},
  \bibinfo{journal}{Phys. Rev. Lett.} \textbf{\bibinfo{volume}{112}},
  \bibinfo{pages}{180603} (\bibinfo{year}{2014}).

\bibitem[{\citenamefont{Ma}(2020)}]{2020-finite-size}
\bibinfo{author}{\bibfnamefont{Y.-H.} \bibnamefont{Ma}},
  \bibinfo{journal}{Entropy} \textbf{\bibinfo{volume}{22}},
  \bibinfo{pages}{1002} (\bibinfo{year}{2020}).

\bibitem[{Del()}]{DeltaI}
\bibinfo{note}{Here, the information correlation $\Delta I$ is related to the
  irreversible entropy generation [37], which is vanish in the reversible
  processes in quasi-static limit.}

\bibitem[{\citenamefont{Landauer}(1961)}]{landauer1961irreversibility}
\bibinfo{author}{\bibfnamefont{R.}~\bibnamefont{Landauer}},
  \bibinfo{journal}{IBM J. Res. Dev.} \textbf{\bibinfo{volume}{5}},
  \bibinfo{pages}{183} (\bibinfo{year}{1961}).

\bibitem[{\citenamefont{Bennett}(1982)}]{bennett1982thermodynamics}
\bibinfo{author}{\bibfnamefont{C.~H.} \bibnamefont{Bennett}},
  \bibinfo{journal}{Int. J. Theor. Phys.} \textbf{\bibinfo{volume}{21}},
  \bibinfo{pages}{905} (\bibinfo{year}{1982}).

\bibitem[{\citenamefont{Dong et~al.}(2011)\citenamefont{Dong, Xu, Cai, Sun
  et~al.}}]{dong2011quantum}
\bibinfo{author}{\bibfnamefont{H.}~\bibnamefont{Dong}},
  \bibinfo{author}{\bibfnamefont{D.}~\bibnamefont{Xu}},
  \bibinfo{author}{\bibfnamefont{C.}~\bibnamefont{Cai}},
  \bibinfo{author}{\bibfnamefont{C.}~\bibnamefont{Sun}}, \bibnamefont{et~al.},
  \bibinfo{journal}{Phys. Rev. E} \textbf{\bibinfo{volume}{83}},
  \bibinfo{pages}{061108} (\bibinfo{year}{2011}).

\bibitem[{\citenamefont{Scully et~al.}(1997)\citenamefont{Scully, Zubairy
  et~al.}}]{scully1997quantum}
\bibinfo{author}{\bibfnamefont{M.~O.} \bibnamefont{Scully}},
  \bibinfo{author}{\bibfnamefont{M.~S.} \bibnamefont{Zubairy}},
  \bibnamefont{et~al.}, \emph{\bibinfo{title}{Quantum Optics}}
  (\bibinfo{publisher}{Cambridge University Press}, \bibinfo{year}{1997}).

\bibitem[{\citenamefont{Napoli et~al.}(2016)\citenamefont{Napoli, Bromley,
  Cianciaruso, Piani, Johnston, and Adesso}}]{napoli2016robustness}
\bibinfo{author}{\bibfnamefont{C.}~\bibnamefont{Napoli}},
  \bibinfo{author}{\bibfnamefont{T.~R.} \bibnamefont{Bromley}},
  \bibinfo{author}{\bibfnamefont{M.}~\bibnamefont{Cianciaruso}},
  \bibinfo{author}{\bibfnamefont{M.}~\bibnamefont{Piani}},
  \bibinfo{author}{\bibfnamefont{N.}~\bibnamefont{Johnston}}, \bibnamefont{and}
  \bibinfo{author}{\bibfnamefont{G.}~\bibnamefont{Adesso}},
  \bibinfo{journal}{Phys. Rev. Lett.} \textbf{\bibinfo{volume}{116}},
  \bibinfo{pages}{150502} (\bibinfo{year}{2016}).

\bibitem[{\citenamefont{Yuan et~al.}(2015)\citenamefont{Yuan, Zhou, Cao, and
  Ma}}]{yuan2015intrinsic}
\bibinfo{author}{\bibfnamefont{X.}~\bibnamefont{Yuan}},
  \bibinfo{author}{\bibfnamefont{H.}~\bibnamefont{Zhou}},
  \bibinfo{author}{\bibfnamefont{Z.}~\bibnamefont{Cao}}, \bibnamefont{and}
  \bibinfo{author}{\bibfnamefont{X.}~\bibnamefont{Ma}}, \bibinfo{journal}{Phys.
  Rev. A} \textbf{\bibinfo{volume}{92}}, \bibinfo{pages}{022124}
  (\bibinfo{year}{2015}).

\bibitem[{\citenamefont{Rossnagel et~al.}(2016)\citenamefont{Rossnagel,
  Dawkins, Tolazzi, Abah, Lutz, Schmidt-Kaler, and Singer}}]{Rossnagel2016}
\bibinfo{author}{\bibfnamefont{J.}~\bibnamefont{Rossnagel}},
  \bibinfo{author}{\bibfnamefont{S.~T.} \bibnamefont{Dawkins}},
  \bibinfo{author}{\bibfnamefont{K.~N.} \bibnamefont{Tolazzi}},
  \bibinfo{author}{\bibfnamefont{O.}~\bibnamefont{Abah}},
  \bibinfo{author}{\bibfnamefont{E.}~\bibnamefont{Lutz}},
  \bibinfo{author}{\bibfnamefont{F.}~\bibnamefont{Schmidt-Kaler}},
  \bibnamefont{and} \bibinfo{author}{\bibfnamefont{K.}~\bibnamefont{Singer}},
  \bibinfo{journal}{Science} \textbf{\bibinfo{volume}{352}},
  \bibinfo{pages}{325} (\bibinfo{year}{2016}).

\bibitem[{\citenamefont{Deng et~al.}(2018)\citenamefont{Deng, Chenu, Diao, Li,
  Yu, Coulamy, del Campo, and Wu}}]{shortcutHEexperimentSciAdv2018}
\bibinfo{author}{\bibfnamefont{S.}~\bibnamefont{Deng}},
  \bibinfo{author}{\bibfnamefont{A.}~\bibnamefont{Chenu}},
  \bibinfo{author}{\bibfnamefont{P.}~\bibnamefont{Diao}},
  \bibinfo{author}{\bibfnamefont{F.}~\bibnamefont{Li}},
  \bibinfo{author}{\bibfnamefont{S.}~\bibnamefont{Yu}},
  \bibinfo{author}{\bibfnamefont{I.}~\bibnamefont{Coulamy}},
  \bibinfo{author}{\bibfnamefont{A.}~\bibnamefont{del Campo}},
  \bibnamefont{and} \bibinfo{author}{\bibfnamefont{H.}~\bibnamefont{Wu}},
  \bibinfo{journal}{Sci. Adv.} \textbf{\bibinfo{volume}{4}},
  \bibinfo{pages}{5909} (\bibinfo{year}{2018}).

\bibitem[{\citenamefont{Bouton et~al.}(2020)\citenamefont{Bouton, Nettersheim,
  Burgardt, Adam, Lutz, and Widera}}]{atomic-Endoreversible2020}
\bibinfo{author}{\bibfnamefont{Q.}~\bibnamefont{Bouton}},
  \bibinfo{author}{\bibfnamefont{J.}~\bibnamefont{Nettersheim}},
  \bibinfo{author}{\bibfnamefont{S.}~\bibnamefont{Burgardt}},
  \bibinfo{author}{\bibfnamefont{D.}~\bibnamefont{Adam}},
  \bibinfo{author}{\bibfnamefont{E.}~\bibnamefont{Lutz}}, \bibnamefont{and}
  \bibinfo{author}{\bibfnamefont{A.}~\bibnamefont{Widera}},
  \bibinfo{journal}{arXiv preprint arXiv:2009.10946}  (\bibinfo{year}{2020}).

\bibitem[{\citenamefont{Holubec and Ryabov}(2017)}]{Holubec2017}
\bibinfo{author}{\bibfnamefont{V.}~\bibnamefont{Holubec}} \bibnamefont{and}
  \bibinfo{author}{\bibfnamefont{A.}~\bibnamefont{Ryabov}},
  \bibinfo{journal}{Phys. Rev. E} \textbf{\bibinfo{volume}{96}}
  (\bibinfo{year}{2017}).

\bibitem[{\citenamefont{Tu}(2021)}]{tu2021abstract}
\bibinfo{author}{\bibfnamefont{Z.-C.} \bibnamefont{Tu}},
  \bibinfo{journal}{Front. Phys.} \textbf{\bibinfo{volume}{16}},
  \bibinfo{pages}{1} (\bibinfo{year}{2021}).

\bibitem[{\citenamefont{Chen et~al.}(2019)\citenamefont{Chen, Sun, and
  Dong}}]{chen2019boosting}
\bibinfo{author}{\bibfnamefont{J.-F.} \bibnamefont{Chen}},
  \bibinfo{author}{\bibfnamefont{C.-P.} \bibnamefont{Sun}}, \bibnamefont{and}
  \bibinfo{author}{\bibfnamefont{H.}~\bibnamefont{Dong}},
  \bibinfo{journal}{Phys. Rev. E} \textbf{\bibinfo{volume}{100}},
  \bibinfo{pages}{032144} (\bibinfo{year}{2019}).

\bibitem[{\citenamefont{Fei et~al.}(2021)\citenamefont{Fei, Chen, and
  Ma}}]{fei2021efficiency}
\bibinfo{author}{\bibfnamefont{Z.}~\bibnamefont{Fei}},
  \bibinfo{author}{\bibfnamefont{J.-F.} \bibnamefont{Chen}}, \bibnamefont{and}
  \bibinfo{author}{\bibfnamefont{Y.-H.} \bibnamefont{Ma}},
  \bibinfo{journal}{arXiv:2109.12816}  (\bibinfo{year}{2021}).

\bibitem[{\citenamefont{Ma et~al.}(2018{\natexlab{b}})\citenamefont{Ma, Xu,
  Dong, and Sun}}]{yhmaoptimalcontrol}
\bibinfo{author}{\bibfnamefont{Y.-H.} \bibnamefont{Ma}},
  \bibinfo{author}{\bibfnamefont{D.}~\bibnamefont{Xu}},
  \bibinfo{author}{\bibfnamefont{H.}~\bibnamefont{Dong}}, \bibnamefont{and}
  \bibinfo{author}{\bibfnamefont{C.-P.} \bibnamefont{Sun}},
  \bibinfo{journal}{Phys. Rev. E} \textbf{\bibinfo{volume}{98}},
  \bibinfo{pages}{022133} (\bibinfo{year}{2018}{\natexlab{b}}).

\bibitem[{\citenamefont{Abiuso and Perarnau-Llobet}(2020)}]{2020Optimal}
\bibinfo{author}{\bibfnamefont{P.}~\bibnamefont{Abiuso}} \bibnamefont{and}
  \bibinfo{author}{\bibfnamefont{M.}~\bibnamefont{Perarnau-Llobet}},
  \bibinfo{journal}{Phys. Rev. Lett.} \textbf{\bibinfo{volume}{124}},
  \bibinfo{pages}{110606} (\bibinfo{year}{2020}).

\bibitem[{\citenamefont{Ma et~al.}(2021)\citenamefont{Ma, Sun, and
  Dong}}]{ma2020self}
\bibinfo{author}{\bibfnamefont{Y.-H.} \bibnamefont{Ma}},
  \bibinfo{author}{\bibfnamefont{C.-P.} \bibnamefont{Sun}}, \bibnamefont{and}
  \bibinfo{author}{\bibfnamefont{H.}~\bibnamefont{Dong}},
  \bibinfo{journal}{Commun. Theor. Phys.
  https://doi.org/10.1088/1572-9494/ac2cb8}  (\bibinfo{year}{2021}).

\bibitem[{\citenamefont{Zhang et~al.}(2009)\citenamefont{Zhang, Cai, You, and
  Zhan}}]{zhang2009hidden}
\bibinfo{author}{\bibfnamefont{B.}~\bibnamefont{Zhang}},
  \bibinfo{author}{\bibfnamefont{Q.-Y.} \bibnamefont{Cai}},
  \bibinfo{author}{\bibfnamefont{L.}~\bibnamefont{You}}, \bibnamefont{and}
  \bibinfo{author}{\bibfnamefont{M.-S.} \bibnamefont{Zhan}},
  \bibinfo{journal}{Phys. Lett. B} \textbf{\bibinfo{volume}{675}},
  \bibinfo{pages}{98} (\bibinfo{year}{2009}).

\bibitem[{\citenamefont{Ma et~al.}(2018{\natexlab{c}})\citenamefont{Ma, Cai,
  Dong, and Sun}}]{ma2018non}
\bibinfo{author}{\bibfnamefont{Y.-H.} \bibnamefont{Ma}},
  \bibinfo{author}{\bibfnamefont{Q.-Y.} \bibnamefont{Cai}},
  \bibinfo{author}{\bibfnamefont{H.}~\bibnamefont{Dong}}, \bibnamefont{and}
  \bibinfo{author}{\bibfnamefont{C.-P.} \bibnamefont{Sun}},
  \bibinfo{journal}{EPL (Europhysics Lett.)} \textbf{\bibinfo{volume}{122}},
  \bibinfo{pages}{30001} (\bibinfo{year}{2018}{\natexlab{c}}).

\bibitem[{\citenamefont{Parikh and Wilczek}(2000)}]{parikh2000hawking}
\bibinfo{author}{\bibfnamefont{M.~K.} \bibnamefont{Parikh}} \bibnamefont{and}
  \bibinfo{author}{\bibfnamefont{F.}~\bibnamefont{Wilczek}},
  \bibinfo{journal}{Phys. Rev. Lett.} \textbf{\bibinfo{volume}{85}},
  \bibinfo{pages}{5042} (\bibinfo{year}{2000}).

\bibitem[{\citenamefont{Xu et~al.}(2014)\citenamefont{Xu, Li, Liu, and
  Sun}}]{xu2014noncanonical}
\bibinfo{author}{\bibfnamefont{D.}~\bibnamefont{Xu}},
  \bibinfo{author}{\bibfnamefont{S.-W.} \bibnamefont{Li}},
  \bibinfo{author}{\bibfnamefont{X.}~\bibnamefont{Liu}}, \bibnamefont{and}
  \bibinfo{author}{\bibfnamefont{C.}~\bibnamefont{Sun}},
  \bibinfo{journal}{Phys. Rev. E} \textbf{\bibinfo{volume}{90}},
  \bibinfo{pages}{062125} (\bibinfo{year}{2014}).

\bibitem[{\citenamefont{Ma et~al.}(2018{\natexlab{d}})\citenamefont{Ma, Chen,
  and Sun}}]{ma2018dark}
\bibinfo{author}{\bibfnamefont{Y.-H.} \bibnamefont{Ma}},
  \bibinfo{author}{\bibfnamefont{J.-F.} \bibnamefont{Chen}}, \bibnamefont{and}
  \bibinfo{author}{\bibfnamefont{C.-P.} \bibnamefont{Sun}},
  \bibinfo{journal}{Nucl. Phys. B} \textbf{\bibinfo{volume}{931}},
  \bibinfo{pages}{418} (\bibinfo{year}{2018}{\natexlab{d}}).

\end{thebibliography}


\begin{thebibliography}{2}
\expandafter\ifx\csname natexlab\endcsname\relax\def\natexlab#1{#1}\fi
\expandafter\ifx\csname bibnamefont\endcsname\relax
  \def\bibnamefont#1{#1}\fi
\expandafter\ifx\csname bibfnamefont\endcsname\relax
  \def\bibfnamefont#1{#1}\fi
\expandafter\ifx\csname citenamefont\endcsname\relax
  \def\citenamefont#1{#1}\fi
\expandafter\ifx\csname url\endcsname\relax
  \def\url#1{\texttt{#1}}\fi
\expandafter\ifx\csname urlprefix\endcsname\relax\def\urlprefix{URL }\fi
\providecommand{\bibinfo}[2]{#2}
\providecommand{\eprint}[2][]{\url{#2}}

\bibitem[{\citenamefont{Baumgratz et~al.}(2014)\citenamefont{Baumgratz, Cramer,
  and Plenio}}]{Quantifying2014}
\bibinfo{author}{\bibfnamefont{T.}~\bibnamefont{Baumgratz}},
  \bibinfo{author}{\bibfnamefont{M.}~\bibnamefont{Cramer}}, \bibnamefont{and}
  \bibinfo{author}{\bibfnamefont{M.}~\bibnamefont{Plenio}},
  \bibinfo{journal}{Phys. Rev. Lett.} \textbf{\bibinfo{volume}{113}},
  \bibinfo{pages}{140401} (\bibinfo{year}{2014}).

\bibitem[{\citenamefont{Scully et~al.}(1997)\citenamefont{Scully, Zubairy
  et~al.}}]{scully1997quantum}
\bibinfo{author}{\bibfnamefont{M.~O.} \bibnamefont{Scully}},
  \bibinfo{author}{\bibfnamefont{M.~S.} \bibnamefont{Zubairy}},
  \bibnamefont{et~al.}, \emph{\bibinfo{title}{Quantum Optics}}
  (\bibinfo{publisher}{Cambridge University Press}, \bibinfo{year}{1997}).

\end{thebibliography}

\end{document}

\title{\textit{\Large{}Supplementary Materials for}{\Large{} \textquotedbl Works
with quantum resource of coherence\textquotedbl}}
\author{Yu-Han Ma}
\address{Graduate School of China Academy of Engineering Physics, No. 10 Xibeiwang
East Road, Haidian District, Beijing, 100193, China}
\author{C. L. Liu}
\address{Graduate School of China Academy of Engineering Physics, No. 10 Xibeiwang
East Road, Haidian District, Beijing, 100193, China}
\author{C. P. Sun}
\email{suncp@gscaep.ac.cn}

\address{Graduate School of China Academy of Engineering Physics, No. 10 Xibeiwang
East Road, Haidian District, Beijing, 100193, China}
\address{Beijing Computational Science Research Center, Beijing 100193, China}

\maketitle
In this Supplementary materials we show the detailed derivation of
Eqs. (7) and (10) of the main text in Sec. \ref{sec:The-entropy-flow}.
In Sec. \ref{sec:Time-evolution-of}, we study the dynamic evolution
of the Jaynes-Cummings model. The change in coherence of all the atoms
during the isothermal process is obtained.

\section{\label{sec:The-entropy-flow}The entropy flow $\dot{S}_{\mathrm{E}}$
of the reservoir}

According to the definition of von-Neumann entropy, the entropy flow
of $\mathrm{E}$ follows

\begin{align}
\dot{S}_{\mathrm{E}} & =\frac{d}{dt}\left\{ -\mathrm{tr}\left[\rho_{\mathrm{E}}\left(t\right)\mathrm{ln}\rho_{\mathrm{E}}\left(t\right)\right]\right\} \\
 & =-\frac{d}{dt}\mathrm{tr}\left[\rho_{\mathrm{E}}\left(t\right)\mathrm{ln}\rho_{\mathrm{E}}\left(t\right)-\rho_{\mathrm{E}}\left(t\right)\mathrm{ln}\rho_{\mathrm{E}}^{\mathrm{eq}}\left(0\right)+\rho_{\mathrm{E}}\left(t\right)\mathrm{ln}\rho_{\mathrm{E}}^{\mathrm{eq}}\left(0\right)\right],\label{eq:Sedot}
\end{align}
where we introduced the initial effective equilibrium density matrix
of the reservoir $\mathrm{E}$ as

\begin{equation}
\rho_{\mathrm{E}}^{\mathrm{eq}}\left(0\right)\equiv\frac{e^{-\beta_{\mathrm{E}}H_{\mathrm{E}}}}{\mathrm{tr}\left(e^{-\beta_{\mathrm{E}}H_{\mathrm{E}}}\right)}\label{eq:ED-E}
\end{equation}
with $\beta_{\mathrm{E}}=\beta_{\mathrm{E}}\left(0\right)$ being
the corresponding effective inverse temperature, which is the solution
of the following equation

\begin{equation}
U_{\mathrm{E}}\left(0\right)=\mathrm{tr}\left[\rho_{\mathrm{E}}\left(0\right)H_{\mathrm{E}}\right]=\frac{\mathrm{tr}\left(e^{-\beta_{\mathrm{E}}H_{\mathrm{E}}}H_{\mathrm{E}}\right)}{\mathrm{tr}\left(e^{-\beta_{\mathrm{E}}H_{\mathrm{E}}}\right)}=-\left\{ \frac{\partial\mathrm{ln}\left[\mathrm{tr}\left(e^{-\beta_{\mathrm{E}}H_{\mathrm{E}}}\right)\right]}{\partial\beta_{\mathrm{E}}}\right\} _{t=0}\label{eq:beta0}
\end{equation}
Substituting Eq. (\ref{eq:ED-E}) into Eq. (\ref{eq:Sedot}), we obtain

\begin{align}
\dot{S}_{\mathrm{E}} & =-\frac{d}{dt}\mathrm{tr}\left[\rho_{\mathrm{E}}\left(t\right)\mathrm{ln}\rho_{\mathrm{E}}\left(t\right)-\rho_{\mathrm{E}}\left(t\right)\mathrm{ln}\rho_{\mathrm{E}}^{\mathrm{eq}}\left(0\right)\right]-\frac{d}{dt}\mathrm{tr}\left[\rho_{\mathrm{E}}\left(t\right)\left(-\beta_{\mathrm{E}}H_{\mathrm{E}}\right)\right]\\
 & =-\dot{S}\left[\rho_{\mathrm{E}}\left(t\right)||\rho_{\mathrm{E}}^{\mathrm{eq}}\left(0\right)\right]+\beta_{\mathrm{E}}\dot{U}_{\mathrm{E}}.\label{eq:Sedot1}
\end{align}
Note that the internal energy change of $\mathrm{E}$ only comes from
the heat exchange between $\mathrm{S}$ and $\mathrm{E}$, such that
$\dot{U}_{\mathrm{E}}=\dot{Q}_{\mathrm{E}}=-\dot{Q}_{\mathrm{S}}$,
then Eq. (\ref{eq:Sedot1}) is further written as

\begin{equation}
\dot{S}_{\mathrm{E}}=-\dot{S}\left[\rho_{\mathrm{E}}\left(t\right)||\rho_{\mathrm{E}}^{\mathrm{eq}}\left(0\right)\right]-\beta_{\mathrm{E}}\dot{Q}_{\mathrm{S}}.
\end{equation}
{[}Eq. (7){]} of the main text is proved. 

We further demonstrate the coherence effect and finite size effect
in $S\left[\rho_{\mathrm{E}}\left(t\right)||\rho_{\mathrm{E}}^{\mathrm{eq}}\left(0\right)\right]$.
According to the definition of quantum relative entropy, $S\left[\rho_{\mathrm{E}}\left(t\right)||\rho_{\mathrm{E}}^{\mathrm{eq}}\left(0\right)\right]$
reads

\begin{equation}
S\left[\rho_{\mathrm{E}}\left(t\right)||\rho_{\mathrm{E}}^{\mathrm{eq}}\left(0\right)\right]=S\left[\rho_{\mathrm{E}}\left(t\right)||\rho_{\mathrm{E}}^{\mathrm{d}}\left(t\right)\right]+S\left[\rho_{\mathrm{E}}^{\mathrm{d}}\left(t\right)||\rho_{\mathrm{E}}^{\mathrm{eq}}\left(0\right)\right],\label{eq:Sroud-roueq}
\end{equation}
where $\rho_{\mathrm{E}}^{\mathrm{d}}\left(t\right)$ is the diagonal
part of $\rho_{\mathrm{E}}\left(t\right)$. On the one hand, the first
term in the right hand of the above equation is 

\begin{align}
S\left[\rho_{\mathrm{E}}\left(t\right)||\rho_{\mathrm{E}}^{\mathrm{d}}\left(t\right)\right] & =\mathrm{tr}\left[\rho_{\mathrm{E}}\left(t\right)\mathrm{ln}\rho_{\mathrm{E}}\left(t\right)-\rho_{\mathrm{E}}\left(t\right)\mathrm{ln}\rho_{\mathrm{E}}^{\mathrm{d}}\left(t\right)\right]\\
 & =\mathrm{tr}\left[\rho_{\mathrm{E}}\left(t\right)\mathrm{ln}\rho_{\mathrm{E}}\left(t\right)-\rho_{\mathrm{E}}^{\mathrm{d}}\left(t\right)\mathrm{ln}\rho_{\mathrm{E}}^{\mathrm{d}}\left(t\right)\right]\\
 & =S\left[\rho_{\mathrm{E}}^{\mathrm{d}}\left(t\right)\right]-S\left[\rho_{\mathrm{E}}\left(t\right)\right]\\
 & =\mathscr{C}_{\mathrm{E}}\left(t\right),\label{eq:QRE}
\end{align}
which is exactly the so-called relative entropy of coherence in quantum
information \citep{Quantifying2014}. 

On the other hand, by straightforward calculation, one obtains the
second term in the right hand of Eq. (\ref{eq:Sroud-roueq}), i.e.,
$S\left[\rho_{\mathrm{E}}^{\mathrm{d}}\left(t\right)||\rho_{\mathrm{E}}^{\mathrm{eq}}\left(0\right)\right]$,
as

\begin{align}
S\left[\rho_{\mathrm{E}}^{\mathrm{d}}\left(t\right)||\rho_{\mathrm{E}}^{\mathrm{eq}}\left(0\right)\right] & =\mathrm{tr}\left[\rho_{\mathrm{E}}^{\mathrm{d}}\left(t\right)\mathrm{ln}\rho_{\mathrm{E}}^{\mathrm{d}}\left(t\right)-\rho_{\mathrm{E}}^{\mathrm{d}}\left(t\right)\mathrm{ln}\rho_{\mathrm{E}}^{\mathrm{eq}}\left(0\right)\right]\\
 & =\mathrm{tr}\left[\rho_{\mathrm{E}}^{\mathrm{d}}\left(t\right)\mathrm{ln}\rho_{\mathrm{E}}^{\mathrm{d}}\left(t\right)-\rho_{\mathrm{E}}^{\mathrm{d}}\left(t\right)\mathrm{ln}\rho_{\mathrm{E}}^{\mathrm{eq}}\left(t\right)+\rho_{\mathrm{E}}^{\mathrm{d}}\left(t\right)\mathrm{ln}\rho_{\mathrm{E}}^{\mathrm{eq}}\left(t\right)-\rho_{\mathrm{E}}^{\mathrm{d}}\left(t\right)\mathrm{ln}\rho_{\mathrm{E}}^{\mathrm{eq}}\left(0\right)\right]\\
 & =S\left[\rho_{\mathrm{E}}^{\mathrm{d}}\left(t\right)||\rho_{\mathrm{E}}^{\mathrm{eq}}\left(t\right)\right]+\mathrm{tr}\left[\rho_{\mathrm{E}}^{\mathrm{d}}\left(t\right)\mathrm{ln}\frac{\rho_{\mathrm{E}}^{\mathrm{eq}}\left(t\right)}{\rho_{\mathrm{E}}^{\mathrm{eq}}\left(0\right)}\right]\\
 & =S\left[\rho_{\mathrm{E}}^{\mathrm{d}}\left(t\right)||\rho_{\mathrm{E}}^{\mathrm{eq}}\left(t\right)\right]+\left\langle \mathrm{ln}\frac{\rho_{\mathrm{E}}^{\mathrm{eq}}\left(t\right)}{\rho_{\mathrm{E}}^{\mathrm{eq}}\left(0\right)}\right\rangle ,\label{eq:SE}
\end{align}
where

\begin{equation}
\rho_{\mathrm{E}}^{\mathrm{eq}}\left(t\right)\equiv\frac{e^{-\beta_{\mathrm{E}}\left(t\right)H_{\mathrm{E}}}}{\mathrm{tr}\left(e^{-\beta_{\mathrm{E}}\left(t\right)H_{\mathrm{E}}}\right)}
\end{equation}
is defined as the effective equilibrium density matrix of $\mathrm{E}$
at time $t$. Similar to Eq. (\ref{eq:beta0}), in the above relation,
$\beta_{\mathrm{E}}\left(t\right)$ is the effective final inverse
temperature of the reservoir, which satisfy

\begin{align}
U_{\mathrm{E}}\left(t\right) & =\mathrm{tr}\left[\rho_{\mathrm{E}}\left(t\right)H_{\mathrm{E}}\right]=-\left\{ \frac{\partial\mathrm{ln}\left[\mathrm{tr}\left(e^{-\beta_{\mathrm{E}}H_{\mathrm{E}}}\right)\right]}{\partial\beta_{\mathrm{E}}}\right\} _{t}\label{eq:}\\
 & =U_{\mathrm{E}}\left(0\right)+\Delta Q_{\mathrm{E}}=U_{\mathrm{E}}\left(0\right)-\Delta Q_{\mathrm{S}}\label{eq:First law for Qs}
\end{align}
The diagonal part $\rho_{\mathrm{E}}^{\mathrm{d}}\left(t\right)$
of the density matrix can be expanded with the energy eigen-state
$\left\{ \left|n\right\rangle \right\} $ of the reservoir as

\begin{equation}
\rho_{\mathrm{E}}^{\mathrm{d}}\left(t\right)=\sum_{n}p_{n}\left(t\right)\left|n\right\rangle \left\langle n\right|.
\end{equation}
And for $\rho_{\mathrm{E}}^{\mathrm{eq}}\left(t\right)$ and $\rho_{\mathrm{E}}^{\mathrm{eq}}\left(0\right)$,
we have

\begin{equation}
p_{n}^{\mathrm{eq}}\left(t\right)=\left\langle n\right|\rho_{\mathrm{E}}^{\mathrm{eq}}\left(t\right)\left|n\right\rangle \equiv\frac{e^{-\beta_{\mathrm{E}}\left(t\right)E_{n}}}{\sum_{n}e^{-\beta_{\mathrm{E}}\left(t\right)E_{n}}}\equiv\frac{e^{-\beta_{\mathrm{E}}\left(t\right)E_{n}}}{Z\left[\beta_{\mathrm{E}}\left(t\right)\right]},
\end{equation}
and

\begin{equation}
p_{n}^{\mathrm{eq}}\left(0\right)=\left\langle n\right|\rho_{\mathrm{E}}^{\mathrm{eq}}\left(0\right)\left|n\right\rangle \equiv\frac{e^{-\beta_{\mathrm{E}}\left(0\right)E_{n}}}{\sum_{n}e^{-\beta_{\mathrm{E}}\left(0\right)E_{n}}}\equiv\frac{e^{-\beta_{\mathrm{E}}\left(0\right)E_{n}}}{Z\left[\beta_{\mathrm{E}}\left(0\right)\right]}.
\end{equation}
Therefore, the second term in the right hand of Eq. (\ref{eq:SE})
becomes 

\begin{align}
\left\langle \mathrm{ln}\frac{\rho_{\mathrm{E}}^{\mathrm{eq}}\left(t\right)}{\rho_{\mathrm{E}}^{\mathrm{eq}}\left(0\right)}\right\rangle  & =\sum_{n}p_{n}\left(t\right)\mathrm{ln}\left\{ \frac{\frac{e^{-\beta_{\mathrm{E}}\left(t\right)E_{n}}}{Z\left[\beta_{\mathrm{E}}\left(t\right)\right]}}{\frac{e^{-\beta_{\mathrm{E}}\left(0\right)E_{n}}}{Z\left[\beta_{\mathrm{E}}\left(0\right)\right]}}\right\} \nonumber \\
 & =\left[\beta_{\mathrm{E}}\left(0\right)-\beta_{\mathrm{E}}\left(t\right)\right]U_{\mathrm{E}}\left(t\right)-\mathrm{ln}\frac{Z\left[\beta_{\mathrm{E}}\left(t\right)\right]}{Z\left[\beta_{\mathrm{E}}\left(0\right)\right]}.\label{eq:rou1-rou2}
\end{align}
For the change in the effective inverse temperature of the reservoir
$\Delta\beta_{\mathrm{E}}$ is small, we have

\begin{equation}
\mathrm{ln}\frac{Z\left[\beta_{\mathrm{E}}\left(t\right)\right]}{Z\left[\beta_{\mathrm{E}}\left(0\right)\right]}=\mathrm{ln}\frac{Z\left[\beta_{\mathrm{E}}\left(0\right)\right]+\left(\frac{\partial Z}{\partial\beta_{\mathrm{E}}}\right)_{t=0}\Delta\beta_{\mathrm{E}}+\frac{1}{2}\left(\frac{\partial^{2}Z}{\partial\beta_{\mathrm{E}}^{2}}\right)_{t=0}\Delta\beta_{\mathrm{E}}^{2}+O\left(\Delta\beta_{\mathrm{E}}^{2}\right)}{Z\left[\beta_{\mathrm{E}}\left(0\right)\right]},\label{eq:LnZ1/Z2}
\end{equation}
with

\begin{equation}
\left(\frac{\partial Z}{\partial\beta_{\mathrm{E}}}\right)_{t=0}=-\sum_{n}e^{-\beta_{\mathrm{E}}\left(0\right)E_{n}}E_{n}=-Z\left[\beta_{\mathrm{E}}\left(0\right)\right]U_{\mathrm{E}}\left(0\right),
\end{equation}
and

\begin{equation}
\left(\frac{\partial^{2}Z}{\partial\beta_{\mathrm{E}}^{2}}\right)_{t=0}=\sum_{n}e^{-\beta_{\mathrm{E}}\left(0\right)E_{n}}E_{n}^{2}=Z\left[\beta_{\mathrm{E}}\left(0\right)\right]\left\langle E_{n}^{2}\right\rangle .
\end{equation}
Here, $\left\langle E_{n}\right\rangle =U_{\mathrm{E}}\left(0\right)=\mathrm{tr}\left[\rho_{\mathrm{E}}^{\mathrm{eq}}\left(0\right)H_{\mathrm{E}}\right]$
and $\left\langle E_{n}^{2}\right\rangle =\mathrm{tr}\left[\rho_{\mathrm{E}}^{\mathrm{eq}}\left(0\right)H_{\mathrm{E}}^{2}\right]$
are respectively the first and second moments of the energy of the
reservoir (average over $\rho_{\mathrm{E}}^{\mathrm{eq}}\left(0\right)$).
By keeping to the second order of $\Delta\beta_{\mathrm{E}}$, Eq.
(\ref{eq:LnZ1/Z2}) is approximated as

\begin{align}
\mathrm{ln}\frac{Z\left[\beta_{\mathrm{E}}\left(t\right)\right]}{Z\left[\beta_{\mathrm{E}}\left(0\right)\right]} & =\mathrm{ln}\left\{ 1-U_{\mathrm{E}}\left(0\right)\Delta\beta_{\mathrm{E}}+\frac{1}{2}\left\langle E_{n}^{2}\right\rangle \Delta\beta_{\mathrm{E}}^{2}+O\left(\Delta\beta_{\mathrm{E}}^{2}\right)\right\} \\
 & =-U_{\mathrm{E}}\left(0\right)\Delta\beta_{\mathrm{E}}+\frac{1}{2}\left\langle E_{n}^{2}\right\rangle \Delta\beta_{\mathrm{E}}^{2}-\frac{1}{2}\left[-U_{\mathrm{E}}\left(0\right)\Delta\beta_{\mathrm{E}}+\frac{1}{2}\left\langle E_{n}^{2}\right\rangle \Delta\beta_{\mathrm{E}}^{2}\right]^{2}+O\left(\Delta\beta_{\mathrm{E}}^{2}\right)\\
 & \approx-U_{\mathrm{E}}\left(0\right)\Delta\beta_{\mathrm{E}}+\frac{1}{2}\Delta\beta_{\mathrm{E}}^{2}\left[\left\langle E_{n}^{2}\right\rangle -U_{\mathrm{E}}^{2}\left(0\right)\right]\\
 & =-U_{\mathrm{E}}\left(0\right)\Delta\beta_{\mathrm{E}}+\frac{1}{2}\Delta\beta_{\mathrm{E}}^{2}\left[\left\langle E_{n}^{2}\right\rangle -\left\langle E_{n}\right\rangle ^{2}\right]\label{eq:Z/Z1}
\end{align}
Substituting Eq. (\ref{eq:Z/Z1}) into Eq. (\ref{eq:rou1-rou2}),
we obtain

\begin{align}
\left\langle \mathrm{ln}\frac{\rho_{\mathrm{E}}^{\mathrm{eq}}\left(t\right)}{\rho_{\mathrm{E}}^{\mathrm{eq}}\left(0\right)}\right\rangle  & =\left[\beta_{\mathrm{E}}\left(0\right)-\beta_{\mathrm{E}}\left(t\right)\right]U_{\mathrm{E}}\left(t\right)-\left\{ -U_{\mathrm{E}}\left(0\right)\Delta\beta_{\mathrm{E}}+\frac{1}{2}\Delta\beta_{\mathrm{E}}^{2}\left[\left\langle E_{n}^{2}\right\rangle -\left\langle E_{n}\right\rangle ^{2}\right]\right\} \\
 & =\Delta\beta_{\mathrm{E}}\left[U_{\mathrm{E}}\left(0\right)-U_{\mathrm{E}}\left(t\right)\right]-\frac{1}{2}\Delta\beta_{\mathrm{E}}^{2}\left(\left\langle E_{n}^{2}\right\rangle -\left\langle E_{n}\right\rangle ^{2}\right)\\
 & =\Delta\beta_{\mathrm{E}}\Delta Q_{\mathrm{S}}-\frac{C_{\mathrm{E}}\Delta\beta_{\mathrm{E}}^{2}}{2\beta_{\mathrm{E}}^{2}},\label{eq:rou1/rou2}
\end{align}
where $C_{\mathrm{E}}=\beta_{\mathrm{E}}^{2}\left(\left\langle E_{n}^{2}\right\rangle -\left\langle E_{n}\right\rangle ^{2}\right)$
is the heat capacity of the reservoir. According to Eq. (\ref{eq:First law for Qs}),
we have

\begin{equation}
\Delta Q_{\mathrm{S}}=-\left[U_{\mathrm{E}}\left(t\right)-U_{\mathrm{E}}\left(0\right)\right]=-\frac{\partial U_{\mathrm{E}}}{\partial T_{\mathrm{E}}}\Delta T_{\mathrm{E}}+O\left(\Delta T_{\mathrm{E}}^{2}\right)=C_{\mathrm{E}}\frac{\Delta\beta_{\mathrm{E}}}{\beta_{\mathrm{E}}^{2}}+O\left(\Delta T_{\mathrm{E}}^{2}\right),
\end{equation}
 Then, substituting the above relation into Eq. (\ref{eq:rou1/rou2}),
one obtains

\begin{equation}
\left\langle \mathrm{ln}\frac{\rho_{\mathrm{E}}^{\mathrm{eq}}\left(t\right)}{\rho_{\mathrm{E}}^{\mathrm{eq}}\left(0\right)}\right\rangle =\frac{C_{\mathrm{E}}\Delta\beta_{\mathrm{E}}^{2}}{2\beta_{\mathrm{E}}^{2}}=\frac{\Delta Q_{\mathrm{S}}^{2}}{2C_{\mathrm{E}}T_{\mathrm{E}}^{2}},
\end{equation}
where only the second order of $\Delta\beta_{\mathrm{E}}$ is kept.
Finally, Eq. (\ref{eq:SE}) is re-written as 

\begin{equation}
S\left[\rho_{\mathrm{E}}^{\mathrm{d}}\left(t\right)||\rho_{\mathrm{E}}^{\mathrm{eq}}\left(0\right)\right]=S\left[\rho_{\mathrm{E}}^{\mathrm{d}}\left(t\right)||\rho_{\mathrm{E}}^{\mathrm{eq}}\left(t\right)\right]+\frac{\Delta Q_{\mathrm{S}}^{2}}{2C_{\mathrm{E}}T_{\mathrm{E}}^{2}}.\label{eq:SEt-0}
\end{equation}
Combining Eq. (\ref{eq:Sroud-roueq}) and Eq. (\ref{eq:SEt-0}), we
have

\begin{equation}
S\left[\rho_{\mathrm{E}}\left(t\right)||\rho_{\mathrm{E}}^{\mathrm{eq}}\left(0\right)\right]=\mathscr{C}_{\mathrm{E}}\left(t\right)+S\left[\rho_{\mathrm{E}}^{\mathrm{d}}\left(t\right)||\rho_{\mathrm{E}}^{\mathrm{eq}}\left(t\right)\right]+\frac{\Delta Q_{\mathrm{S}}^{2}}{2C_{\mathrm{E}}T_{\mathrm{E}}^{2}}.
\end{equation}
Arrive here, {[}Eq. (10){]} of the main text is proved.

\section{\label{sec:Time-evolution-of}Time evolution of the Jaynes-Cummings
model}

In this section, we specific the general isolated system considered
in the main text as a single mode Jaynes-Cummings (JC) model, where
we denote the two-level atom as reservoir, and the light field as
the system. 

Let $\omega_{a}$ and $\omega$ being the frequency of the two-level
atom and the light field respectively, and denote $\left|e\right\rangle $
($\left|g\right\rangle $ ) as the excited (ground) state of the atom,
$\left\{ \left|n\right\rangle \right\} $ as the particle number states
of the light field, the Hamiltonian of the JC model reads, taking
$\hbar=1$,

\begin{equation}
H=\omega_{a}\left|e\right\rangle \left\langle e\right|+\omega a^{\dagger}a+g\left(a^{\dagger}\sigma_{-}+a\sigma_{+}\right),
\end{equation}
where $\sigma_{+}=\left(\sigma_{-}\right)^{\dagger}=\left|e\right\rangle \left\langle g\right|$,
$a$ is the annihilation operator of the light field, and $g$ is
the coupling constant. The eigen-states of the JC model, in the subspace
$\left\{ \left|e,n\right\rangle ,\left|g,n+1\right\rangle \right\} $,
are \citep{scully1997quantum}

\begin{align}
\left|+,n\right\rangle  & =\cos\frac{\theta_{n}}{2}\left|e,n\right\rangle +\sin\frac{\theta_{n}}{2}\left|g,n+1\right\rangle ,\label{eq:eigen1}\\
\left|-,n\right\rangle  & =\sin\frac{\theta_{n}}{2}\left|e,n\right\rangle -\cos\frac{\theta_{n}}{2}\left|g,n+1\right\rangle ,\label{eq:eigen2}
\end{align}
respectively, and the corresponding eigen-energies are

\begin{equation}
E_{n}^{\pm}=\left(n+\frac{1}{2}\right)\omega+\frac{1}{2}\omega_{a}\pm\sqrt{\left(n+1\right)g^{2}+\frac{\delta^{2}}{4}}
\end{equation}
with $\delta=\omega_{a}-\omega$, and 

\begin{equation}
\tan\frac{\theta_{n}}{2}=\frac{2\sqrt{n+1}g}{\sqrt{4\left(n+1\right)g^{2}+\delta^{2}}}.
\end{equation}

To study the time evolution of the JC model, we first assume the initial
density matrix of the whole system follows

\begin{equation}
\rho\left(0\right)=\rho_{a}\left(0\right)\otimes\rho_{l}\left(0\right)=\left(p_{e}\left|e\right\rangle \left\langle e\right|+p_{g}\left|g\right\rangle \left\langle g\right|+\mu\left|e\right\rangle \left\langle g\right|+\mu^{*}\left|g\right\rangle \left\langle e\right|\right)\otimes\left(\sum_{n}\frac{e^{-\beta n\omega}}{Z}\left|n\right\rangle \left\langle n\right|\right),
\end{equation}
where $\rho_{a}\left(0\right)$ and $\rho_{l}\left(0\right)$ are
respectively the initial density matrix of the atom and the light
field, and we have assumed the light field is in a thermal equilibrium
state with temperature $T=1/\left(k_{\mathrm{B}}\beta\right).$ $p_{e}$
and $p_{g}=1-p_{e}$ are respectively the population of the excited
state and ground state. $\mu$ is non-diagonal element of $\rho_{a}\left(0\right)$,
which describes the coherence of the atom. The time evolution of the
total system is

\begin{equation}
\rho\left(t\right)=U\left(t\right)\rho\left(0\right)U^{\dagger}(t)=e^{-iHt}\rho\left(0\right)e^{iHt}.
\end{equation}
For the state $\left|e,n\right\rangle $, one has

\begin{align*}
\left|e,n\right\rangle _{t} & \equiv U\left(t\right)\left|e,n\right\rangle =e^{-iHt}\left(\cos\frac{\theta_{n}}{2}\left|+,n\right\rangle +\sin\frac{\theta_{n}}{2}\left|-,n\right\rangle \right)\\
 & =e^{-iE_{n}^{+}t}\cos\frac{\theta_{n}}{2}\left|+,n\right\rangle +e^{-iE_{n}^{-}t}\sin\frac{\theta_{n}}{2}\left|-,n\right\rangle \\
 & =e^{-iE_{n}^{+}t}\cos\frac{\theta_{n}}{2}\left(\cos\frac{\theta_{n}}{2}\left|e,n\right\rangle +\sin\frac{\theta_{n}}{2}\left|g,n+1\right\rangle \right)+e^{-iE_{n}^{-}t}\sin\frac{\theta_{n}}{2}\left(\sin\frac{\theta_{n}}{2}\left|e,n\right\rangle -\cos\frac{\theta_{n}}{2}\left|g,n+1\right\rangle \right)\\
 & =\left(e^{-iE_{n}^{+}t}\cos^{2}\frac{\theta_{n}}{2}+e^{-iE_{n}^{-}t}\sin^{2}\frac{\theta_{n}}{2}\right)\left|e,n\right\rangle +\frac{1}{2}\sin\theta_{n}\left(e^{-iE_{n}^{+}t}-e^{-iE_{n}^{-}t}\right)\left|g,n+1\right\rangle .
\end{align*}
Similarly, for $\left|g,n\right\rangle $

\begin{align*}
\left|g,n\right\rangle _{t} & \equiv U\left(t\right)\left|g,n\right\rangle =e^{-iHt}\left(\sin\frac{\theta_{n-1}}{2}\left|+,n-1\right\rangle -\cos\frac{\theta_{n-1}}{2}\left|-,n-1\right\rangle \right)\\
 & =e^{-iE_{n-1}^{+}t}\sin\frac{\theta_{n-1}}{2}\left|+,n-1\right\rangle -e^{-iE_{n-1}^{-}t}\cos\frac{\theta_{n-1}}{2}\left|-,n-1\right\rangle \\
 & =e^{-iE_{n-1}^{+}t}\sin\frac{\theta_{n-1}}{2}\left(\cos\frac{\theta_{n-1}}{2}\left|e,n-1\right\rangle +\sin\frac{\theta_{n-1}}{2}\left|g,n\right\rangle \right)-e^{-iE_{n-1}^{-}t}\cos\frac{\theta_{n-1}}{2}\left(\sin\frac{\theta_{n-1}}{2}\left|e,n-1\right\rangle -\cos\frac{\theta_{n-1}}{2}\left|g,n\right\rangle \right)\\
 & =\left(e^{-iE_{n-1}^{+}t}\sin^{2}\frac{\theta_{n-1}}{2}+e^{-iE_{n-1}^{-}t}\cos^{2}\frac{\theta_{n-1}}{2}\right)\left|g,n\right\rangle +\frac{1}{2}\sin\theta_{n-1}\left(e^{-iE_{n-1}^{+}t}-e^{-iE_{n-1}^{-}t}\right)\left|e,n-1\right\rangle ,
\end{align*}
The density matrix of the whole system at time $t$ can be expressed
with $\left|g,n\right\rangle _{t}$ and $\left|e,n\right\rangle _{t}$
as

\begin{equation}
\rho\left(t\right)=\sum_{n}\frac{e^{-\beta n\omega}}{Z}\left(p_{e}\left|e,n\right\rangle _{t}\left\langle e,n\right|+p_{g}\left|g,n\right\rangle _{t}\left\langle g,n\right|+\mu\left|e,n\right\rangle _{t}\left\langle g,n\right|+\mu^{*}\left|g,n\right\rangle _{t}\left\langle e,n\right|\right).
\end{equation}
Then, the reduced density matrix of the atom reads

\begin{align}
\rho_{a}\left(t\right) & =\textrm{Tr}_{l}\left[\rho\left(t\right)\right]=\sum_{n'}\left\langle n'\right|\rho\left(t\right)\left|n'\right\rangle \\
 & =\sum_{n}\frac{e^{-\beta n\omega}}{Z}\sum_{n'}\left\langle n'\right|\left(p_{e}\left|e,n\right\rangle _{t}\left\langle e,n\right|+p_{g}\left|g,n\right\rangle _{t}\left\langle g,n\right|+\mu\left|e,n\right\rangle _{t}\left\langle g,n\right|+\mu^{*}\left|g,n\right\rangle _{t}\left\langle e,n\right|\right)\left|n'\right\rangle ,
\end{align}
where, by straightforward calculation, one has

\begin{equation}
\left\langle n'\right|\left.e,n\right\rangle _{t}=\left(e^{-iE_{n}^{+}t}\cos^{2}\frac{\theta_{n}}{2}+e^{-iE_{n}^{-}t}\sin^{2}\frac{\theta_{n}}{2}\right)\left|e\right\rangle \delta_{n'n}+\frac{1}{2}\sin\theta_{n}\left(e^{-iE_{n}^{+}t}-e^{-iE_{n}^{-}t}\right)\left|g\right\rangle \delta_{n',n+1},
\end{equation}
and

\begin{equation}
\left\langle n'\right|\left.g,n\right\rangle _{t}=\left(e^{-iE_{n-1}^{+}t}\sin^{2}\frac{\theta_{n-1}}{2}+e^{-iE_{n-1}^{-}t}\cos^{2}\frac{\theta_{n-1}}{2}\right)\left|g\right\rangle \delta_{n'n}+\frac{1}{2}\sin\theta_{n-1}\left(e^{-iE_{n-1}^{+}t}-e^{-iE_{n-1}^{-}t}\right)\left|e\right\rangle \delta_{n',n-1}
\end{equation}
The reduced density matrix of the atom is obtained as
\begin{equation}
\rho_{a}\left(t\right)=p_{e}\left(t\right)\left|e\right\rangle \left\langle e\right|+\left[1-p_{e}\left(t\right)\right]\left|g\right\rangle \left\langle g\right|+\mu\left(t\right)\left|e\right\rangle \left\langle g\right|+\mu^{*}\left(t\right)\left|g\right\rangle \left\langle e\right|,
\end{equation}
where, by defining $\Delta E_{n}\equiv E_{n}^{+}-E_{n}^{-}=\sqrt{4\left(n+1\right)g^{2}+\delta^{2}}$,

\begin{align}
p_{e}\left(t\right) & =\sum_{n}\frac{e^{-\beta n\omega}}{Z}\left\{ p_{e}\left|e^{-iE_{n}^{+}t}\cos^{2}\frac{\theta_{n}}{2}+e^{-iE_{n}^{-}t}\sin^{2}\frac{\theta_{n}}{2}\right|^{2}+p_{g}\left|\frac{1}{2}\sin\theta_{n-1}\left(e^{-iE_{n-1}^{+}t}-e^{-iE_{n-1}^{-}t}\right)\right|^{2}\right\} \\
 & =\sum_{n}\frac{e^{-\beta n\omega}}{Z}\left\{ p_{e}\left[1-\frac{1}{2}\sin^{2}\theta_{n}\left(1-\cos\Delta E_{n}t\right)\right]+p_{g}\frac{1}{2}\sin^{2}\theta_{n-1}\left(1-\cos\Delta E_{n-1}t\right)\right\} 
\end{align}
and

\begin{equation}
\mu\left(t\right)=\mu\sum_{n}\frac{e^{-\beta n\omega}}{Z}\left(e^{-iE_{n}^{+}t}\cos^{2}\frac{\theta_{n}}{2}+e^{-iE_{n}^{-}t}\sin^{2}\frac{\theta_{n}}{2}\right)\left(e^{-iE_{n-1}^{+}t}\sin^{2}\frac{\theta_{n-1}}{2}+e^{-iE_{n-1}^{-}t}\cos^{2}\frac{\theta_{n-1}}{2}\right)^{*}
\end{equation}
For simplicity, we focus on the resonance case with $\delta=0$, in
this situation

\begin{equation}
E_{n}^{\pm}=\left(n+1\right)\omega\pm g\sqrt{\left(n+1\right)},\tan\frac{\theta_{n}}{2}=\frac{2\sqrt{n+1}g}{\sqrt{4\left(n+1\right)g^{2}+\delta^{2}}}=1,
\end{equation}
and $p_{e}\left(t\right)$ and $\mu\left(t\right)$ can be further
simplified as 

\begin{equation}
p_{e}\left(t\right)=\frac{p_{e}}{2}+\frac{1}{4}-\frac{p_{g}}{4Z}+\left(\frac{1}{4}p_{e}-\frac{e^{-\beta\omega}}{4}p_{g}\right)\sum_{n=0}\frac{e^{-\beta n\omega}}{Z}\cos\left(2g\sqrt{n+1}\right)t,\label{eq:pet}
\end{equation}
and

\begin{equation}
\eta\left(t\right)=\mu e^{-i\omega t}\sum_{n}\frac{e^{-\beta n\omega}}{Z}\cos\left(\sqrt{n+1}gt\right)\cos\left(\sqrt{n}gt\right)\label{eq:etat}
\end{equation}
respectively. In the short time limit of $gt\ll1$, by keeping the
second order of $gt$, Eq. (\ref{eq:pet}) and Eq. (\ref{eq:etat})
are approximated as

\begin{equation}
p_{e}\left(t\right)=p_{e}-\frac{g^{2}t^{2}}{2\left(1-e^{-\beta\omega}\right)}\left(p_{e}-e^{-\beta\omega}p_{g}\right),
\end{equation}
and

\begin{equation}
\mu\left(t\right)=\mu e^{-i\omega t}\left[1-\left(\frac{1}{2}+\left\langle n\right\rangle \right)g^{2}t^{2}\right],
\end{equation}
with $\left\langle n\right\rangle =\sum_{n}ne^{-\beta n\omega}/Z$
the initial mean particle number of the light field. The relative
entropy of coherence of the atom at time $t$ is, according to the
definition,

\begin{align}
\xi_{a}\left(t\right) & \equiv\mathrm{tr}\left[\rho_{a}\left(t\right)\ln\rho_{a}\left(t\right)\right]-\mathrm{tr}\left[\rho_{a}^{d}\left(t\right)\ln\rho_{a}^{d}\left(t\right)\right],\\
 & =-S\left[\rho_{a}\left(t\right)\right]+S\left[\rho_{a}^{d}\left(t\right)\right]
\end{align}
where

\begin{equation}
\rho_{a}^{d}\left(t\right)=p_{e}\left(t\right)\left|e\right\rangle \left\langle e\right|+p_{g}\left(t\right)\left|g\right\rangle \left\langle g\right|.
\end{equation}
Note that, the eigen-values of $\rho_{a}\left(t\right)$ are

\begin{equation}
p_{\pm}\left(t\right)=\frac{1}{2}\pm\sqrt{\left|\mu\left(t\right)\right|^{2}+\left[\frac{1}{2}-p_{e}\left(t\right)\right]^{2}},
\end{equation}
then $S\left[\rho_{a}\left(t\right)\right]$ and $S\left[\rho_{a}^{d}\left(t\right)\right]$
are explicitly written as

\begin{equation}
S\left[\rho_{a}\left(t\right)\right]=-\left[1-p_{+}\left(t\right)\right]\mathrm{ln}\left[1-p_{+}\left(t\right)\right]-p_{+}\left(t\right)\mathrm{ln}p_{+}\left(t\right),
\end{equation}
and

\begin{equation}
S\left[\rho_{a}^{d}\left(t\right)\right]=-\left[1-p_{e}\left(t\right)\right]\mathrm{ln}\left[1-p_{e}\left(t\right)\right]-p_{e}\left(t\right)\mathrm{ln}p_{e}\left(t\right).
\end{equation}
We further assume the coherence of the atom is week, namely, $\left|\mu\left(t\right)\right|\ll\frac{1}{2}-p_{e}\left(t\right)$,
such that

\begin{equation}
p_{+}\left(t\right)\approx1-p_{e}\left(t\right)+\frac{\left|\mu\left(t\right)\right|^{2}}{1-2p_{e}\left(t\right)},
\end{equation}
where the higher orders of $\left|\mu\left(t\right)\right|^{2}$ are
ignored. Then the relative entropy of coherence is approximated as,
by keeping the first order of $\left|\mu\left(t\right)\right|^{2}$,

\begin{align}
\xi_{a}\left(t\right) & =-\left.\frac{\partial S\left[\rho_{a}\left(t\right)\right]}{\partial p_{+}\left(t\right)}\right|_{\mu\left(t\right)=0}\left\{ p_{+}\left(t\right)-\left[1-p_{e}\left(t\right)\right]\right\} \\
 & =\mathrm{ln}\frac{p_{g}\left(t\right)}{1-p_{g}\left(t\right)}\frac{\left|\mu\left(t\right)\right|^{2}}{1-2p_{e}\left(t\right)}.
\end{align}
Finally, the change in relative entropy of coherence in a short time
interval $\tau\ll g^{-1}$ is obtained as

\begin{align}
\Delta\xi_{a}\left(\tau\right) & =\xi_{a}\left(\tau\right)-\xi_{a}\left(0\right)\\
 & =\frac{1}{1-2p_{e}}\mathrm{ln}\frac{p_{g}}{1-p_{g}}\left\{ \mu^{2}\left[1-\left(\frac{1}{2}+\left\langle n\right\rangle \right)g^{2}\tau^{2}\right]^{2}-\mu^{2}\right\} \\
 & =-\frac{\mu^{2}}{1-2p_{e}}\mathrm{ln}\frac{p_{g}}{1-p_{g}}\left(1+2\left\langle n\right\rangle \right)g^{2}\tau^{2}\\
 & =-\xi_{a}\left(0\right)\left(1+2\left\langle n\right\rangle \right)g^{2}\tau^{2}
\end{align}
Here, only the second order of $g\tau$ is kept. For the $i$-th atom
intercting with the photon field, the change in its relative entropy
of coherence is $\xi_{a}^{(i)}\left(\tau\right)=\Delta\xi_{a}\left(\tau\right)$.
After $N$ atoms pass though the light field with the rate $r$ and
operation time $t_{\mathrm{f}}=N\tau=Nr^{-1}$, the change in relative
entropy of coherence of all $N$ the atoms is

\begin{equation}
\Delta\xi_{\mathrm{E}}\left(t_{\mathrm{f}}\right)=\sum_{i=1}^{i=N}\Delta\xi_{a}^{(i)}\left(\tau\right)=N\Delta\xi_{a}\left(\tau\right)=-\varGamma\xi_{a}\left(0\right)t_{\mathrm{f}}
\end{equation}
where 

\begin{equation}
\varGamma=g^{2}r^{-1}\left(1+2\left\langle n\right\rangle \right),
\end{equation}
is an increasing function of the coupling constant $g$ and the mean
particle number $\left\langle n\right\rangle $, and an decreasing
function of the passing rate $r$ of the atoms.

\bibliographystyle{apsrev}
\addcontentsline{toc}{section}{\refname}\bibliography{QRCT}